\documentclass[fleqn,usenatbib]{mnras}

\usepackage{graphicx}
\usepackage{newtxtext,newtxmath}
\usepackage[T1]{fontenc}
\usepackage{ae,aecompl}
\usepackage{adjustbox}
\usepackage{amsfonts}
\usepackage{amsmath}
\usepackage{graphicx}	
\usepackage{tabularx}
\usepackage{verbatim}
\usepackage{xspace}
\usepackage{adjustbox}
\usepackage{rotating, lipsum, babel}
\usepackage{xparse}

\NewDocumentCommand{\Msun}{o}{
  \IfNoValueTF{#1}
    {\,\mathrm{M}_{\odot}}
    {10^{#1}\,\mathrm{M}_{\odot}}
}

\newcommand{\pc}{\,\mathrm{pc}}
\newcommand{\kpc}{\,\mathrm{kpc}}
\newcommand{\Gyr}{\,\mathrm{Gyr}}
\newcommand{\Gyrs}{\,\mathrm{Gyrs}}
\newcommand{\kms}{\,\mathrm{km\,s}^{-1}}

\newcommand{\Rtidal}{R_\mathrm{tidal}}
\newcommand{\ac}{a_\mathrm{c}}

\newcommand{\Rap}{R_\mathrm{ap}}

\newcommand{\mudelta}{\mu_\delta}
\newcommand{\mualpha}{\mu_\alpha}
\newcommand{\mualphac}{\mualpha\cos{\delta}}
\newcommand{\nhat}{\vec{n}}
\newcommand{\Gaia}{\textit{Gaia}\xspace}
\newcommand{\masyr}{\mathrm{mas\,yr}^{-1}}

\newcommand{\VarF}{\mathcal{F}}
\newcommand{\QDSG}{Q_\mathrm{DSG}}
\newcommand{\FDSG}{\VarF_\mathrm{DSG}(p)}
\newcommand{\FDSGE}{\VarF_\mathrm{DSG}(e)}
\newcommand{\FDSGR}{\VarF_\mathrm{DSG}(\Rap)}
\newcommand{\FDSGTheta}{\VarF_\mathrm{DSG}(\Delta \theta)}

\newcommand{\FSgr}{\VarF_\mathrm{Sgr}(p)}
\newcommand{\FSgrE}{\VarF_\mathrm{Sgr}(e)}
\newcommand{\FSgrR}{\VarF_\mathrm{Sgr}(\Rap)}
\newcommand{\FSgrTheta}{\VarF_\mathrm{Sgr}(\Delta \theta)}

\newcommand{\RapGC}{\Rap(\mathrm{GC})}
\newcommand{\RapDSG}{\Rap(\mathrm{DSG})}
\newcommand{\RapSgr}{\Rap(\mathrm{Sgr})}

\newcommand{\pDSG}{p_\mathrm{DSG}}
\newcommand{\pGC}{p_\mathrm{GC}}

\newcommand{\rsc}{r_\mathrm{sc}}
\newcommand{\rc}{r_\mathrm{c}}

\newcommand{\Rh}{R_\mathrm{h}}

\newcommand{\Rc}{R_\mathrm{c}}
\newcommand{\rvir}{r_\mathrm{vir}}
\newcommand{\Dh}{D_\odot}
\newcommand{\aCM}{a_\mathrm{CM}}
\newcommand{\eCM}{e_\mathrm{CM}}

\newcommand{\Mb}{M_\mathrm{b}}
\newcommand{\Md}{M_\mathrm{d}}

\newcommand{\Msgr}{M_\mathrm{Sgr}}
\newcommand{\Mvir}{M_\mathrm{vir}}
\newcommand{\MSgr}{M_{\text{Sgr}}}

\newcommand{\phid}{\phi_\mathrm{disc}}
\newcommand{\phih}{\phi_\mathrm{halo}}

\newcommand{\phisgr}{\phi_\mathrm{Sgr}}

\newcommand{\Vinfty}{V_\infty}

\newcommand{\VLOS}{V_\mathrm{LOS}}

\newcommand{\Nbody}{$N$-body\xspace}

\newcommand{\secref}[1]{Section~\ref{#1}}
\newcommand{\figref}[1]{Figure~\ref{#1}}
\newcommand{\tabref}[1]{Table~\ref{#1}}
\newcommand{\equref}[1]{Equation~\eqref{#1}}
\newcommand{\percent}{\,\mathrm{per}\,\mathrm{cent}}

\defcitealias{Rostami2022}{Paper~I}

\title[The in-situ vs. ex-situ origin of the MW GCs]{Identifying the possible ex-situ origin of the globular clusters of the Milky Way: A kinematic study}

\author[Rostami Shirazi et al.]{
Ali Rostami Shirazi,$^{1}$\thanks{E-mail: a.rostami@iasbs.ac.ir}
Pouria Khalaj,$^{1}$ and
Hosein Haghi,$^{1}$
\\
$^{1}$Department of Physics, Institute for Advanced Studies in Basic Sciences (IASBS), PO Box 11365-9161, Zanjan, Iran\\
}

\date{Accepted XXX. Received YYY; in original form ZZZ}

\pubyear{2023}

\begin{document}
\label{firstpage}
\pagerange{\pageref{firstpage}--\pageref{lastpage}}
\maketitle

\begin{abstract}
This is the second paper in a series, which studies the likelihood that some globular clusters (GCs) of the Milky Way (MW) could have originated from a dwarf satellite galaxy (DSG). Using a large suite of three-body simulations we determine the present-day orbital properties of 154 GCs that could have escaped from 41 MW DSGs over the past $8\Gyrs$. For the MW we considered two sets of static and dynamic models which account for the sustained growth of the MW since its birth. We focus on the Magellanic Clouds and Sagittarius. We compare the apogalactic distance, eccentricity, and orbital inclination of the MW GCs with those of runaway GCs from DSGs, to constrain their possible ex-situ origin. We observe a positive correlation between a DSG mass and the dispersion of its runaway GCs in the orbital parameter space of ($\Rap$, $e$). We provide tables of the identified MW GCs and their likely associated progenitors. In total, we find 29 (19\%) MW GCs which could be kinematically associated with MW DSGs. We report, for the first time, 6 and 10 new associations with the Large Magellanic Cloud and the Sagittarius, respectively. For the Sagittarius we predict a concentration of runaway GCs at large apogalactic distances of $\Rap\approx275-375\kpc$, $e\approx0.8$, and a relative inclination of $\Delta\theta\approx20\degr$. So far, there has not been any observed GCs with such orbital elements. Complemented with photometric and spectroscopic observations, and cosmological simulations, the findings from the present study could conclusively settle the debate over the in-situ vs. ex-situ origin of the MW GCs.
\end{abstract} 

\begin{keywords}
methods: numerical --  globular clusters: general -- galaxies: individual: Sagittarius.
\end{keywords}

\section{Introduction}\label{sec:intro}

According to the standard $\Lambda$ cold dark matter cosmology, all galaxies are thought to grow in a hierarchical fashion in which larger structures are formed through the continuous merging of smaller structures \citep{White1978}. If a dwarf satellite galaxy (DSG) merges with a massive galaxy, its stellar populations are expected to be incorporated into the merged system.  Cosmological simulations indicate that star formation can be tagged as occurring within the potential well radius of the host primary galaxy (i.e. in-situ) or within the satellite galaxy that is eventually accreted (i.e. ex-situ). This suggests that today's galaxies should contain contributions from both in-situ and ex-situ formed globular clusters (GCs) depending on the galaxy's assembly history \citep{Oser2010, Pillepich2015}.

More than 150 GCs have been identified around the Milky Way (MW), whose origin is still an issue. There is observational evidence to imply that the MW has been probably made up of mergers with smaller DSGs so that GCs can be considered as potential tracers of this process  \citep{Searle1978}. A number of GCs in the MW, especially the so-called `young halo' population, are hypothesized to have been captured from DSGs \citep{LAW2010b, Mackey2004}. The ex-situ GCs were born in DSGs and then captured from their host DSG by the MW. The origin of ex-situ GCs can be broadly divided into two categories, 1) GCs that were born in DSGs and their host galaxy is completely merged into the MW; and 2) GCs that originated from a DSG that is orbiting around MW.

\par Up until today several studies have attempted to associate some of MW GCs with known merging process of DSGs. The Sagittarius dwarf spheroidal galaxy (Sgr) is the ﬁrst discovered candidate of a merger with the MW \citep{Ibata1994}. Several studies over the past two decades have associated a number of GCs to the Sgr. A number of studies used photometric data and compared the stellar properties of GCs with those of the Sgr stars, such properties include metallicity, chemical abundances, colour–magnitude diagram and age of stars \citep{Carraro2009,Carraro2011,Carretta2014,Sbordone2015,Carretta2017,Bellazzini2002,Caffau2005,Sbordone2005}. Other studies utilized numerical models that accurately reproduce the position and radial velocity of stars belonging to the Sgr streams, and identified the GCs that may be associated with the Sgr \citep{LAW2010b,belazini2020}. \citet{massari2019} obtained the energy and angular momentum range for GCs that were likely to be associated with the Sgr. The GCs which are thought to be associated with the Sgr are Whiting 1, NGC~6715 (M54), Ter~7, Arp~2, Ter~8, Pal~12, NGC~6284, NGC~5053, NGC~4147, NGC~5634, NGC~5824, Pal~2, and NGC~2419 \citep{LAW2010b,massari2019,belazini2020}. Among these GCs, M~54, Arp~2, Ter~7, and Ter~8 are still bound to the Sgr \citep{belazini2020}.

\citet{kipelman2019} identified new members of the Helmi streams \citep{Helmi1999}, using the complete phase-space information combining \Gaia Data Release 2 (DR2), and the APOGEE DR2, RAVE DR5, and LAMOST DR4 spectroscopic surveys. Moreover, they performed \Nbody simulations to limit the progenitor's properties and timing of accretion. They then labeled GCs that show overlap in energy and angular momentum space with the stream members and concluded that seven GCs are possible candidates to be associated with the Helmi streams (NGC~4590, NGC~5024, NGC~5053, NGC~5272, NGC~5634, NGC~5904, and NGC~6981).

\par The \Gaia-Enceladus-Sausage is an elongated structure in velocity space discovered by \citet{Belokurov2018}. using the kinematics of metal-rich halo stars. They showed that it could be created by a massive dwarf galaxy with a total mass of $5\times\Msun[10]$ on a strongly radial orbit that merged with the MW. \citet{Myeong1018} sought evidence for the associated \Gaia-Enceladus-Sausage GCs by examining the 91 MW GCs' structural data in action space using the \Gaia DR2 catalogue, accompanied with the proper motions obtained from the Hubble Space Telescope proper. They identiﬁed eight GCs as belonging to the \Gaia-Enceladus-Sausage (NGC~1851, NGC~1904, NGC~2298, NGC~2808, NGC~5286, NGC~6864, NGC~6779, and NGC~7089).

\par \citet{massari2019} studied the origin of 151 GCs of the MW. They concluded that only 40\% of the clusters probably formed in-situ, while 35\% of the GCs are possibly associated with known merging processes including the \Gaia-Enceladus-Sausage (19\%), the Sagittarius dwarf galaxy (5\%), the progenitor of the Helmi streams (6\%), and to the Sequoia galaxy (5\%). They did not find any origin for other GCs, of these, about $16\percent$ of the GCs were classified in the high energy group And the rest of them low energy group.

Using numerical simulations, \citet{Khalaj2015, Khalaj2016} studied the escape of GC stars from the Fornax DSG, in the context of multiple-stellar populations in GCs. They showed that Fornax could have lost a substantial amount of its stellar mass as a result of the MW tidal field and gas expulsion occurring inside GCs.
 
\par \citet{Rostami2022}, hereafter \citetalias{Rostami2022}, determined the escape fraction of GCs from 13 massive DSGs of the MW. They demonstrated that the escape fractions are not negligible, i.e. for most DSGs this value was at least $\sim20\percent$ and for two of them it was above $80\percent$. They concluded that it is very likely that a number of MW GCs originated from DSGs. Given the number of GCs observed in the Fornax, Sgr, Small Magellanic Cloud (SMC), and the Large Magellanic Cloud (LMC), and their corresponding escape fractions obtained by \citetalias{Rostami2022}, they estimated that at least two GCs to have escaped from the Fornax, two GCs from the SMC, four GCs from the LMC, and about 14 GCs have escaped from the Sgr. Moreover, they found that the average escape time of GCs from DSGs reaches a plateau at $t\approx8\Gyrs$ which means that the MW DSGs with an age greater than $8\Gyrs$, are not likely to lose any more GCs.

\par The present paper is a continuation of \citetalias{Rostami2022}. Here, we utilize a method based on the kinematic properties of the MW GCs and the MW DSGs. We classify the orbits of the MW GCs and calculate the probability that these GCs are associated with DSGs. Our method does not take into account the photometric/spectroscopic  properties of GCs and DSGs. We focus only on the second category of the origin of ex-situ GCs, i.e. GCs that were born in a DSG that is orbiting around MW, and then escaped from it as a result of the MW tidal field. In particular, we focus on the Sgr, LMC, and the SMC, as the heaviest MW satellite galaxies. We also briefly touch upon merging processes and dissolved DSGs. In the future study, we exclusively investigate the association of the MW GCs with merging processes whose progenitor DSGs are completely merged into the MW.

\par The present paper is divided into four sections. \secref{sec:methodology} includes the description of our methodology and adopted simulation models. In \secref{sec:results} we present our results regarding the association of the MW GCs with some DSGs as well as drawing a comparison with other relevant studies. Finally, in \secref{sec:conclusion}, we summarize the main outcomes of our work, and propose a direction for future studies.

\section{Methodology}\label{sec:methodology}
The principal problem we need to solve is obtaining the orbital parameters of GCs that were initially bound to a given DSG and then escaped from it due to the Galactic tidal field. One can model DSGs assuming they are composed of a stellar system of $N$ stars and a number of bound GCs, whose initial conditions are random, albeit in equilibrium. Such an approach requires solving the \Nbody problem, i.e. calculating the forces that each individual field star exerts on a GC and then summing over all the forces, for a large number of time steps over a period of $\sim8\Gyrs$. This is computationally expensive, hence not well-fit for our purpose of sweeping a large parameter space. 

DSGs are collisionless systems. As a result, one can instead assume a smooth potential field for them. This is an advantage of collisionless systems, allowing us to reduce the \Nbody problem to a three-body problem consisting of the MW, a DSG, and a GC. The usefulness and prominence of such an approach have been previously shown in studies such as \citet{Khalaj2016}. In \secref{sec:Nbody-N3body}, we will show that the tail stream distribution of a DSG, which is obtained by our three-body method is in a good agreement with those obtained by more time-consuming direct \Nbody method. 

This is also the same approach we followed in \citetalias{Rostami2022}. We place the MW at the centre of a right-handed Cartesian coordinate system, whose $X$-axis points towards the location of the Sun, the $Y$-axis is in the direction of the Galaxy's rotation, and the $Z$-axis is determined by the right-hand rule. The MW is assumed to remain still throughout the simulation owing to its large mass. The trajectory of the DSG is only determined by the MW. We take GCs as point masses whose motion is prescribed by both the MW and the DSG. GCs are spatially distributed according to the density profile of their host DSG. The initial velocities of GCs are drawn from a three-dimensional Maxwell-Boltzmann distribution. We make sure that the GCs are initially bound to their host DSGs. GCs that remain within $2\times\Rtidal$ for about four times their orbital period are considered as initially-bound and their trajectory is followed for a simulation time, using a 10th order Runge-Kutta integrator. At the end of each simulation, we designate those GCs whose final distances exceed $2\times\Rtidal$ from the centre of their host DSG. We consider such GCs as runaways. It should be mentioned  that in all simulations we assume that the distribution of GCs follows the distribution of baryonic matter of DSGs. In \secref{sec:filter_sgr}, we will discuss how the results change if any other initial GC distribution is adopted. Note that only the dark halo component is included in the equation of motion of test particles as it is the most dominant component in dwarf galaxies. There are more (minor) details about the initial conditions of the GCs and the method we use to solve the three-body problem. However, for the sake of brevity, we refrain from repeating them here. One can refer to \citetalias{Rostami2022} for a detailed description of our methodology and the relevant initial conditions.

We perform the simulations for a large ensemble of GCs, i.e. until we end up with 2500 runaway GCs for each DSG and obtain their orbital parameters. Then we compare the orbital parameters of the MW GCs with those of runaway GCs of the given DSG and calculate the probability of their association with the given DSG. Moreover, we examine the effect of the DSG mass on the scattering of GCs in the parameter space, using a semi-analytical method.

\subsection{Models}
\subsubsection{The potential field of the MW}\label{sec:MWpot}
For the MW, we adopt the well-known \texttt{MWPotential2014} model, described in \cite{bovy14}. In this model, the MW potential field consists of three components, namely bulge, disc, and halo. The bulge density follows a power-law distribution (spherical) with an exponential cutoff given by
\begin{equation}\label{eq:MW_bulge}
    \rho(r) \propto \frac{1}{r^\alpha} \exp{\left(-\left(\frac{r}{\rc}\right)^2\right)}
\end{equation}
The disc is a \citet{Miyamoto1975} model
\begin{equation}\label{eq:miyamoto}
	\phid = {\frac{-G\Md}{\sqrt{x^2 + y^2 + \left(a+\sqrt{b^2 + z^2}\right)^2}} }
\end{equation}
It also assumes a dark-matter halo with a Navarro–Frenk–White potential (NFW) from \citet{NFW}
\begin{equation}\label{eq:NFW}
\phih = -\frac{G \Mvir}{r} \frac{\log \left(\frac{c\, r}{\rvir}+1\right)}{\log (c+1)-\frac{c}{c+1}}
\end{equation}

It is evident that \texttt{MWPotential2014} is a static model, i.e. the potential is time-invariant. In addition to this model, we also consider a dynamic model for the MW potential which is time-dependent. This model is motivated by the fact that the mass of the MW had been smaller at birth and has continuously grown since then as a result of e.g. merging processes. To model the sustained growth of the MW over the past $8\Gyrs$, we assume that at each point in time, the MW conforms to a potential profile similar to \texttt{MWPotential2014}, where its parameters (masses and scale lengths) are time-dependent. Following \citet{haghi_dympot}, we can derive the required relations for the masses and scale lengths as a function of time. In particular, the virial mass is given by
\begin{equation}\label{eq:exp_M}
\Mvir(z)=\Mvir(0)\exp{\left(-2 z \ac\right)}
\end{equation}
where $z$ is the cosmological redshift and $\ac=0.34$. The virial mass at $z=0$ (present time) is denoted by $\Mvir(0)$, and its value is given by \texttt{MWPotential2014}. The values of $\rvir$ and $c$ vary as a function of time as given by \citet{haghi_dympot}, i.e. equations 9 and 6 therein. The relations for the bulge and disc mass as well as their scale lengths are given by five similar equations which can be expressed in a compact mathematical form as follows
\begin{equation}\label{eq:dyn_bul_dis}
M_{\{\rm{d,b}\}}(z)=\Mvir(z) \frac{M_{\{\rm{d,b}\}}(0)}{\Mvir(0)}
\end{equation}
\begin{equation}\label{eq:dyn_scal}
\left\{a,b,\rc\right\}(z)=\rvir(z) \frac{\left\{a,b,\rc\right\}(0)}{\rvir(0)}
\end{equation}
where the $\{...\}$ notation factors out the variable parts of each equation, while keeping in the identical parts of equations to avoid repetition. \tabref{tab:MWpot} summarizes the values of the aforementioned parameters (masses and scale lengths) for two epochs of $t=0\Gyr$ (present time) and $t=-8\Gyr$.

Similar to \citet{fritz_2018}, we consider two values for the halo mass of the MW, i.e $\Mvir=0.8\times\Msun[12]$ (from \texttt{MWPotential2014}), and $\Mvir=1.6\times\Msun[12]$. All other parameters of the MW potential model remain unchanged. Models with light and heavy halos are denoted by L and H, respectively.

\begin{table}
	\centering
	\begin{tabular}{cccc}
		\hline
		parameter & unit & $t=0\Gyr$ & $t=-8\Gyr$  \\ 
		\hline
		$\alpha$ & 1 & -1.8 & NA\\
		$\rc$ & $\kpc$ & 1.9 & 0.83 \\
		$a$ & $\kpc$ & 3 & 1.32 \\
		$b$ & $\kpc$ & 0.28 & 0.12 \\
		$c$ & $\kpc$ & 15.3 & 7.37 \\
		$\rvir$ & $\kpc$ & 245 & 108 \\
		$\Mvir$ & $\Msun[10]$ & 80 & 38.52 \\		
		$\Mb$ & $\Msun[10]$ & 0.5 & 0.24 \\ 
		$\Md$ & $\Msun[10]$ & 6.8 & 3.27 \\ 
		\hline
	\end{tabular}
	\caption{The parameters of the MW components at the present time ($t=0$) and $8\Gyrs$ ago. Masses and distances are expressed in $\Msun[10]$ and $\kpc$, respectively}
	\label{tab:MWpot}
\end{table}

\begin{table}
	\centering
	\begin{tabular}{ccc}
		\hline
		Model & Potential of the MW & Dynamical Friction?\\ 
		\hline
         L1 & Static & No \\
         L2 & Static & Yes \\
         L3 & Dynamic & No \\
         L4 & Dynamic & Yes \\
         H1 & Static & No \\
         H2 & Static & Yes \\
         H3 & Dynamic & No \\
         H4 & Dynamic & Yes \\
		\hline
	\end{tabular}
	\caption{The characteristics of the simulation models. Models designated as L and H correspond to light and heavy MW halos, respectively. L1, L2, H1, and H2 models have static potentials for both the MW and the Sgr. In comparison, L3, L4, H3, and H4 have dynamic potentials. Models ending in even numbers have dynamical friction, whereas odd-numbered models are free from the effect of dynamical friction. H4 and L4 models are more realistic among all models.}
	\label{tab:sim_model}
\end{table}

\subsubsection{The potential field of the Sgr}\label{sec:sgrpot}
We assume a \citet{Plummer1911} model for the present-day potential of the Sgr as follows
\begin{equation}\label{eq:sgr_plum}
\phisgr(r)=-\frac{G \Msgr}{\sqrt{r^2+\rsc^2}}
\end{equation}

To determine the values of $\Msgr$ and $\rsc$, we use the results of \citet{sgr_pot}. They studied the three-dimensional structure of the Sgr using the astrometric and photometric data of \Gaia. They found that the total mass of the Sgr enclosed within a radius of $5\kpc$ is $4\times\Msun[8]$; and that the peak value of the circular speed of the Sgr is $21\kms$ which is reached at a radius of $\sim2.5-3\kpc$. These findings result in $\Msgr=4.8\times\Msun[8]$ and $\rsc=1.8\kpc$. The half-mass radius ($\Rh$) in a Plummer model relates to $\rsc$ via a linear equation, i.e. $\Rh=1.304\rsc$. This yields $\Rh=2.34\kpc$ for the Sgr. 

\subsubsection{The time-dependent potential models of the Sgr and the corresponding orbits}\label{sec:sim_model}
\par We get the present-day equatorial coordinates ($\alpha, \delta$), proper motions $(\mualphac, \mudelta)$, the line-of-sight velocity ($\VLOS$), and the heliocentric distance ($\Dh$) of the Sgr from the \Gaia DR2 (see \tabref{tab:MW_DSGs_params}). For each of the potential models adopted for the MW, i.e. dynamic and static models as defined in \secref{sec:MWpot}, we trace back the orbit of the Sgr for $8\Gyrs$ to obtain its initial conditions. 

\par In total, we consider eight different models to obtain the orbit of the Sgr. These models are labeled as L1 to L4 (light) and H1 to H4 (heavy). The characteristics of these models are summarized in \tabref{tab:sim_model}.

\par Models with a static potential for the MW and the Sgr are designated by L1 and H1. The models labeled as L2 and H2 are as same as L1/H1, except that the dynamical friction also enters the equations of motion. We apply the standard form of dynamical friction to the halo (e.g. \citealt{Binney2011})

\begin{equation} \label{eq:DYfric}
   F_{\rm{DF}}=-0.428\frac{GM^2}{r^2}\ln{(\Lambda)},
\end{equation}
where $r$ is the distance of the DSG from the centre of the MW, $M$ is the mass of the DSG, and $\ln(\Lambda)=3$ is the Coulomb logarithm. 

\par
\begin{figure}
  \centering
  \includegraphics[width=\linewidth]{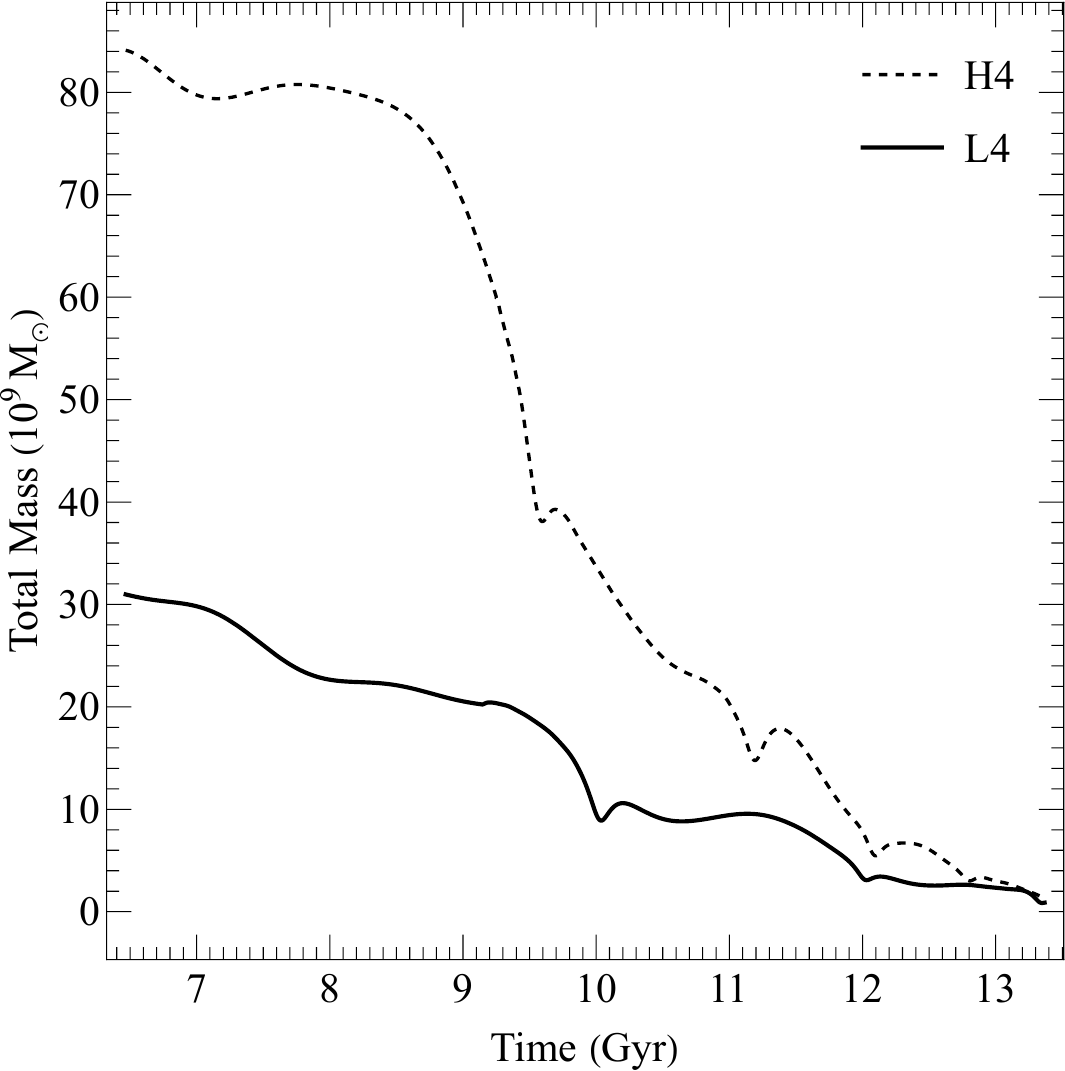}
  \caption{Time evolution of the Sgr total mass within the tidal radius in H4 and L4 models.}
  \label{fig:Sgr-Mt}
\end{figure}

 In the H3 and L3 models, the MW and the Sgr potentials are considered as being dynamic (see \secref{sec:MWpot}). In H4 and L4 models, the Sgr and MW potentials are dynamic and the effect of dynamical friction has been considered. Tidal stripping induced by the MW has reduced the total mass (dark + baryonic matter) of the Sgr over the past $8\Gyrs$. We have quantified this effect using an \Nbody simulation, where the Sgr is made up of $N=50,000$ equal-mass particles distributed according to the Plummer model. The particles do not undergo stellar evolution. The motion of each particle is determined by the combined potential of all other particles as well as the MW, using an 8th-order Runge-Kutta integrator. We then determine the values of $\MSgr(t=-8\Gyr)$ and $\rsc(t=-8\Gyr)$ in such a way that after $8\Gyrs$ of evolution, they match those of the present-day Sgr. In particular, we obtain an initial dark halo mass of $\MSgr(t=-8\Gyr)=6\times\Msun[10]$ for L3, and $\MSgr(t=-8\Gyr)=2.8\times\Msun[11]$ for H3. To obtain the initial conditions of the H4 and L4 models, we consider dynamical friction as well. We find $\MSgr(t=-8\Gyr)=4\times\Msun[10]$ for L4, and $\MSgr(t=-8\Gyr)=1\times\Msun[11]$ for H4, which is in agreement with the results of \citet{Gibbons2017} and \citet{Jiang2000}. Time evolution of the Sgr total mass within the tidal radius in H4 and L4 models is shown in \figref{fig:Sgr-Mt}. Note that the initial stellar mass of Sgr is estimated to be $\sim 10^9 \Msun$ (e.g. \citealt{Niederste-Ostholt2010,Law2010a}), which is smaller than the dark halo mass by one to two orders of magnitude. As a result, we ignore the effect of stellar-mass in the equations of motion.

\figref{fig:Sgr_orbit} shows the orbit of the Sgr for the aforementioned eight models for the past $8\Gyrs$. In H3, H4, L3, and L4 the position of \Nbody particles which constitute the Sgr is known as a function of time. This enables us to calculate the Sgr potential as a function of time and position $\phi(\vec{r},t)$. This time-dependent potential combined with the MW potential is used to determine the trajectory of GCs. 

\subsubsection{Orbital parameters of GCs}
Among all the properties which describe the orbit of a GC we choose the 3-tuple of ($\Rap,e,\nhat$), where $\Rap$ is the apogalactic distance, $e$ is the eccentricity, and $\nhat$ is normal to the orbital plane. Instead of $\nhat$, one can equivalently use the angular momentum vector $\vec{L}$, since $\vec{L}=L\hat{n}$. Within the context of this study, these are sufficient to determine the momentum and energy state of each GC. However, one should note that this 3-tuple does not uniquely determine the orbit of each GC. Moreover, for GCs whose orbital plane as well as the orbital inclination changes, we use the average of $\vec{n}$ vectors over the last $1\Gyr$ of the simulation.

\subsection{Probability of GCs association with the DSGs}\label{sec:prob_sgr}
Using the data from \Gaia DR2 \citep{baumgardt}, we can calculate the position and velocity vectors of MW GCs at the present time in our coordinate system. As a result, the orbital parameters of GCs ($\Rap,e,\nhat$) can be fully obtained. The uncertainties in proper motions $(\mualphac, \mudelta)$ and line-of-sight velocity ($\VLOS$) of GCs and DSGs are usually not negligible and yield large uncertainties in their orbits. In comparison, errors in $\alpha, \delta$, and $\Dh$ are negligible. 

\par To examine the correlation of orbital parameters of MW GCs with the runaway GCs of a DSG, we proceed as follows. First, for each of the observed parameters (proper motions and the line-of-sight velocity) of MW GCs, we consider a set $S(p)$ of three data points, which includes the mean value of the parameter along with its corresponding lower and upper bounds, i.e.
\begin{equation}
    S(p) = \{p-\epsilon_p, p, p+\epsilon_p\}
\end{equation}
where $p\in\{\mualphac, \mudelta, \VLOS\}$ and $\epsilon_p$ is the corresponding error of each parameter. For each MW GC, this gives us a set $S$ of $3^3 = 27$ elements, each element being a different 3-tuple of observed data points. In other words, $S = \prod_{p} S(p)$, where $\prod$ denotes the three-fold Cartesian product of sets since we have three observed parameters. This further translates into a set of 27 3-tuples of orbital parameters ($\Rap,e,\nhat$) for each GC. It is clear that if the error of data is small, these 27 different orbits of a GC are very close. 

\par Next, depending on whether the DSG is the Sgr, the LMC, the SMC, or else we follow different procedures described in one of the following subsections. 

\subsubsection{Association with the Sgr}\label{sec:sgr-assoc}
The orbital parameters of the Sgr are well-measured, hence the corresponding  errors are relatively negligible. As a result, in all simulations, we only use the mean values for its orbital parameters without considering the errors. 
\par
\begin{figure}
  \centering
  \includegraphics[width=\linewidth]{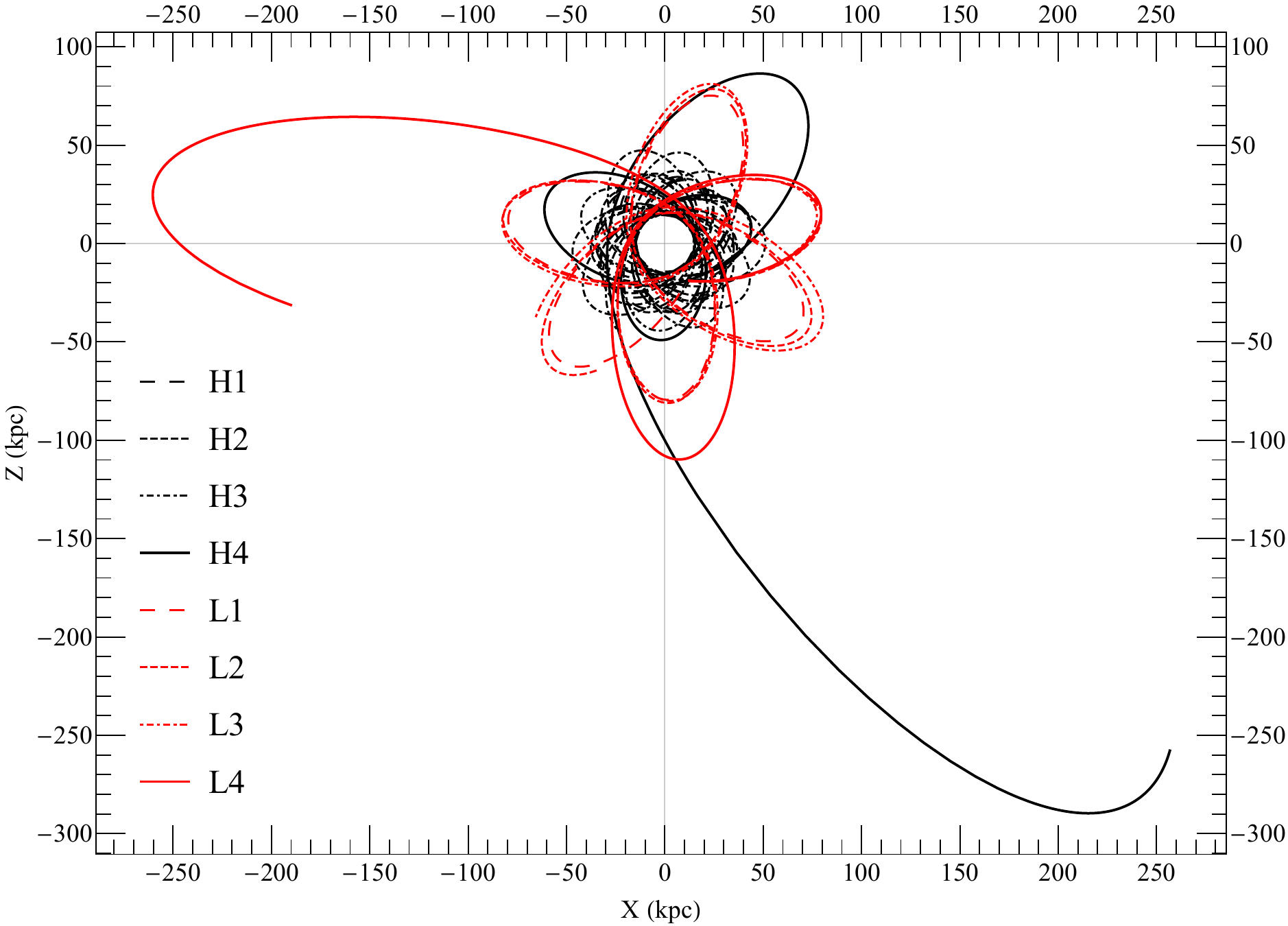}
  \caption{The orbit of the Sgr DSG in our eight simulation models for a period of $8\Gyrs$. Orbits corresponding to light and heavy halo masses are shown in red and black, respectively. The orbits are depicted for the $X-Z$ plane of the Galactocentric coordinate.}
  \label{fig:Sgr_orbit}
\end{figure}

\begin{figure*}
  \centering
  \includegraphics[width=0.74\linewidth]{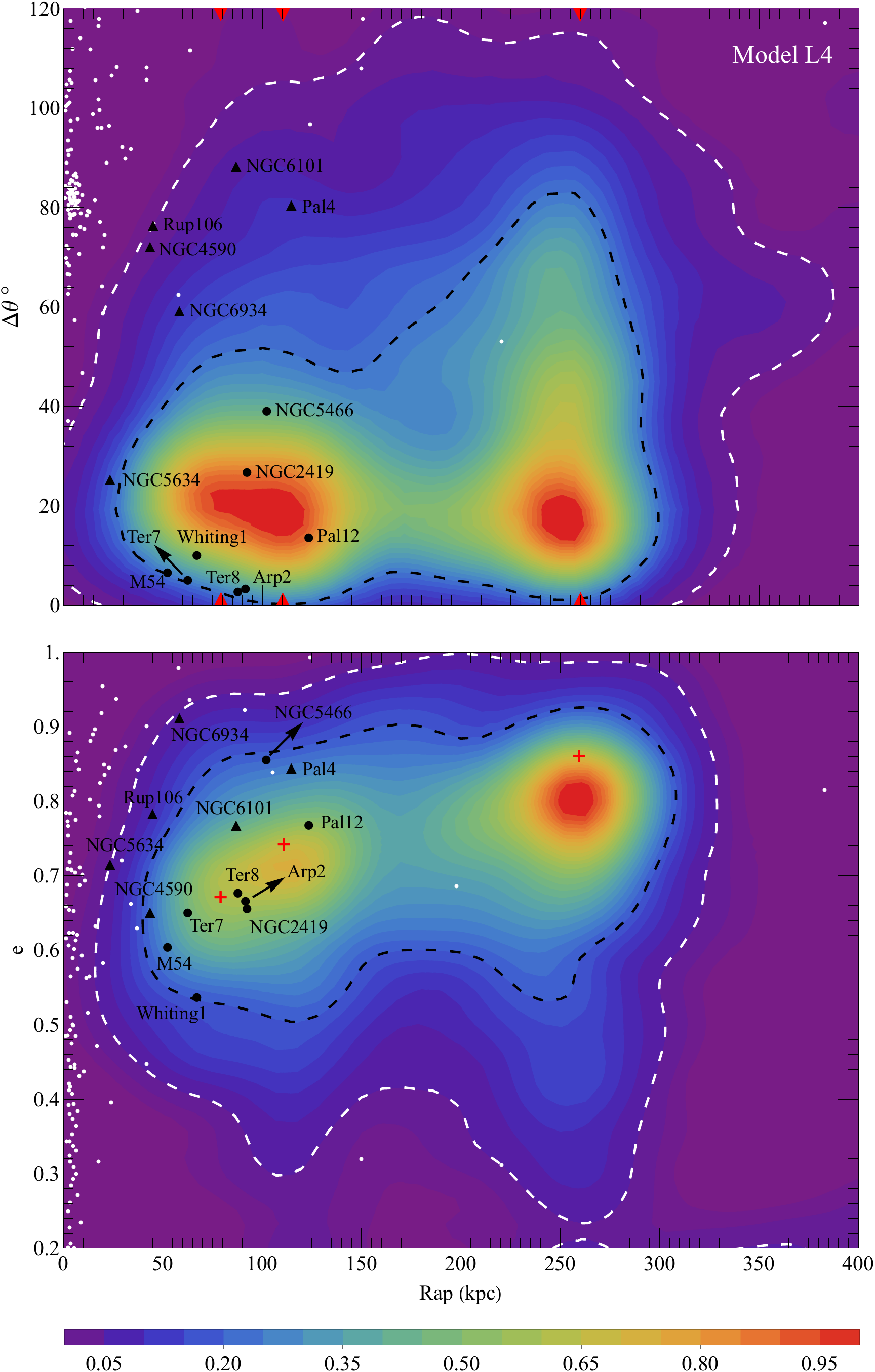}
  \caption{The scaled PDF of the Sgr runaway GCs, colour coded with respect to isodensity contours. The black and white contours correspond to boundaries enclosing $65\%$, and between $65\%$ and $95\%$ of all data, respectively. These provide the basis for categorizing the MW GCs into Flag~1 (black circles), Flag~2 (black triangles), and non-associations (white circles). In the bottom panel, the red crosses mark the location of the Sgr in the parameter space as the Sgr spirals inwards as a result of dynamical friction. Likewise, the red arrowheads mark the apogalactic distances of the Sgr in the top panel, where we have $\Delta\theta=0\degr$ for the Sgr. One can interpret non-associations as GCs with an association probability of less than $5\%$. The apparent placement of some points (e.g. non-associations) within the white or black contours is due to the fact that panels display projections. As a result, one needs to consider both panels together in order to infer the association categories. The elongated peak on the left consists of two sub-peaks, i.e. there exists three peaks in unison with the $\sim3$ apogalactic passages of the Sgr around the MW (see \figref{fig:Sgr_orbit}).}
  \label{fig:pdf-GCs-dis}
\end{figure*}

 We construct a three-dimensional parameter space whose components are $\Rap(\kpc)$, $e$, and $\Delta\theta^{\circ}$, where $\Delta\theta$ is the orbital inclination of a GC measured with respect to the Sgr orbital plane. Each of the Sgr runaway GCs and the MW GCs, occupy a single point in this parameter space. We utilize the multivariate (three-dimensional) probability density function (PDF) of runaway GCs in the aforementioned parameter space as the basis to quantify associations. We employed (Gaussian) kernel density estimation to derive the PDF. \figref{fig:pdf-GCs-dis} illustrates the scaled PDF for the Sgr L4 model, showing the isodensity contours. There seems to exist two peaks in the PDF, where one has a spherical-like form (right peak), while the other is more elongated (left peak). Upon close inspection, it becomes evident that the elongated peak is a blend of two sub-peaks. These three peaks correspond to regions of the highest concentration of runaway GCs in the parameter space and coincide with the apogalactic distance of the Sgr as it completes $\sim3$ orbits around the MW and spirals inwards due to dynamical friction over the past $8\Gyrs$ (ref. \figref{fig:Sgr_orbit}). As we move away from the peaks, the PDF asymptotically reaches zero, indicating regions where the probability of association is extremely low. To consider the errors into account, we first calculate the PDF values for all elements in the associated set $S$ that we have for each MW GC. We then average over these PDF values to obtain a single PDF value for each MW GC. Based on the PDF values we then classify MW GCs into three categories, namely Flag~1, Flag~2, and non-associations. Flag~1 corresponds to MW GCs with a very high association probability with the Sgr. In comparison, Flag~2 GCs have a lower association with the Sgr. These respectively refer to all MW GCs which lie within a boundary (contour), encompassing $65\%$ of all runaway GCs for Flag~1, and between $65\%$ and $95\%$ for Flag~2. Non-associated GCs are highly unlikely to have originated from the Sgr as they lie outside the $95\%$ boundary, i.e. where the probability of association is less than $5\%$.

\subsubsection{Association with the LMC and the SMC}
The LMC is the most massive DSG of the MW. So far, 19 GCs have been observed in the LMC \citep{LAW2010b}. According to \citetalias{Rostami2022}, it is estimated that at least $16\percent$ of LMC GCs should have been stripped off by the MW, with an average escape time of $\approx8\Gyrs$. This implies that the escape process of the LMC GCs has been completed and we don't expect any more runaway GCs in the future. These results translate into four runaway GCs for the LMC. Assuming that the runaway GCs have not yet been dissolved, they should be still orbiting the MW.

\par To investigate the association between the MW GCs and the LMC, we repeat the same procedure as we did for the Sgr. We only consider H1 (without the dynamical friction) and H2 (with the dynamical friction) models. We assume a Plummer density profile for the LMC and omitting the (negligible) errors on its orbital parameters. According to the \citet{bekki2009}, the total dynamical mass and scale length of the LMC are $2\times\Msun[10]$ and 3$\kpc$, respectively. Two more simulations are performed for SMC in H2 model. One with the presence of the LMC and the other without it. Also for the SMC, we assume a Plummer model with a total dynamical mass of $3\times\Msun[9]$ and a scale length of $2\kpc$ \citep{bekki2009}. 

However, it should be mentioned that since the LMC and SMC are irregular galaxies, considering Plummer density profile does not necessarily match the observed profiles. We adopted the Plummer model for the density profile of DSGs as it is a good approximation for spherical objects such as globular clusters and galactic dark matter halo. In \secref{sec:analytical_met}, we will show that the total mass of DSG is the most effective parameter in the distribution of runaway GCs rather than the exact density profile of the host DSGs.

\subsubsection{Association with other MW DSGs}\label{sec:prob_DSG}
The errors on the orbital parameters and masses of other MW DSGs are not necessarily negligible (see \tabref{tab:MW_DSGs_params}). In addition, performing simulations for the other 38 MW DSGs is computationally expensive. As a result, the method we used for the Sgr, the LMC, and the SMC cannot be readily applied to other DSGs. For them, we proceed as follows. First, for each DSG and the set of its runaway GCs ($\QDSG$), we define the filter of an orbital parameter $\FDSG$ as the maximum difference between the value of that parameter for the DSG ($\pDSG$) and that of its runaway GCs ($\pGC$), i.e. 
\begin{equation}
    \FDSG  = \max{\bigg\{\big| \pDSG - \pGC\big| : \forall \ \mathrm{GC} \in \QDSG \bigg\}}
\end{equation}
where $p\in\{\Rap, e, \Delta\theta\}$, and $\Delta\theta$ is the orbital inclination of the runaway GC with respect to the orbital plane of its host DSG. The value of $\FDSG$ gives us a measure of the parameter dispersion in the parameter space for the given DSG. We will demonstrate in \secref{sec:eff_mass} that $\FDSG$ is positively correlated with the mass of the DSG. This indicates that for lighter DSGs, the orbit of runaway GCs is very similar to that of the DSG. After the LMC and the SMC, the Sgr is the heaviest DSG of the MW. The values of $\FDSG$ obtained for these three DSGs can be considered as an upper limit on $\FDSG$ for all other (lighter) DSGs. Among them, the Sgr provides the most stringent bounds. Therefore, we pick $\FSgr$ as a measure to compare the orbit of MW GCs and runaway GCs of a DSG. This alleviates the need for performing more simulations similar to those of \secref{sec:sim_model}.

\par Having found an estimation on $\FDSG$ for all MW DSGs, we check if both of the following conditions are satisfied for all pairs of a MW GC and a DSG 
\begin{equation}\label{eq:condition}
\begin{split}
i) \quad  & \forall p\in\{e, \Delta \theta\} \Rightarrow \big|\pGC - \pDSG\big| \leq \FSgr \\
ii) \quad & \left|1-\frac{\RapGC}{\RapDSG}\right| \leq 
\frac{\FSgrR}{\RapSgr},
\end{split}
\end{equation}
where we would like to reemphasize that GCs refer to MW GCs and not the runaway GCs of DSGs. We will show in \secref{sec:eff_ra}, if the mass and the orbital eccentricity of a DSG is fixed and it is placed in an orbit with larger values for $\Rap$, the corresponding $\FDSGE$ and $\FDSGTheta$ will not change. However, $\FDSGR$ grows as $\RapDSG$ increases, hence condition $ii$ in \equref{eq:condition} is defined.

As mentioned earlier, unlike the Sgr and the LMC, the uncertainties on the orbital parameters of other DSGs are not negligible and cannot be simply omitted. Similar to the set $S$ of 27 initial conditions that we made for each GC, we generate 27 initial conditions for each DSG. This translates into $27\times27=729$ sets of orbital parameters for each pair of a MW GC and a DSG, for each of which we check if the conditions given by \equref{eq:condition} hold. 

\par Finally, for each pair of a MW GC and a DSG, we define the association probability as follows 
\begin{equation} \label{eq:prob_asso}
   \mathcal{P} = \frac{N_\mathrm{true}}{729}
\end{equation}
where $N_\mathrm{true}$ is the number of elements in the set of all orbital parameters for which \equref{eq:condition} holds and $729$ is the total number of elements. Based on this probability we assign association flags to GCs and DSGs, as we did in \secref{sec:prob_sgr}. In particular, a probability of $\mathcal{P}\geq0.6$ is designated as Flag 1 indicating a high association probability, and Flag 2 with $0.2\leq \mathcal{P}<0.6$ corresponds to a lower association probability.

\subsection{Distribution of runaway GCs in the parameter space: a semi-analytical approach}\label{sec:analytical_met}
\subsubsection{dichotomy in the semi-major axes of runaway GCs}\label{sec:dichotomy}

\begin{figure}
  \centering
  \includegraphics[width=1\linewidth]{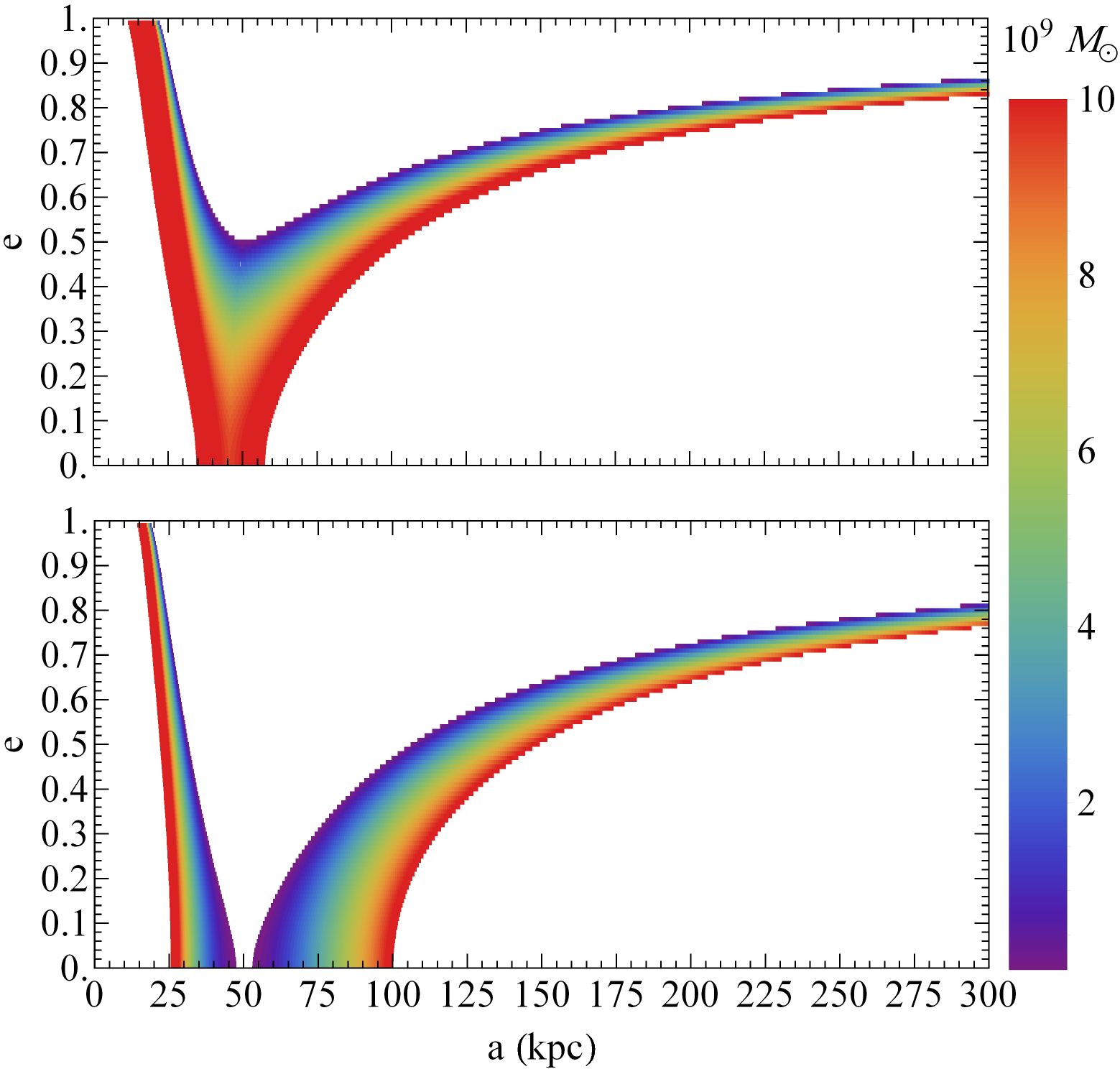}
  \caption{The semi-major axis ($a_2$) of GCs around the MW versus their orbital eccentricities ($e_2$) in the range of $10^8$ to $\Msun[10]$ for $m_1$. The DSG-GC system initially orbits the MW on an orbit with $\aCM=50\kpc$ and eccentricities of $\eCM=0$ (bottom panel) and $\eCM=0.5$ (top panel).}
  \label{fig:e_vs_a}
\end{figure}

\begin{figure*}
    \centering
    \includegraphics[width=1\linewidth]{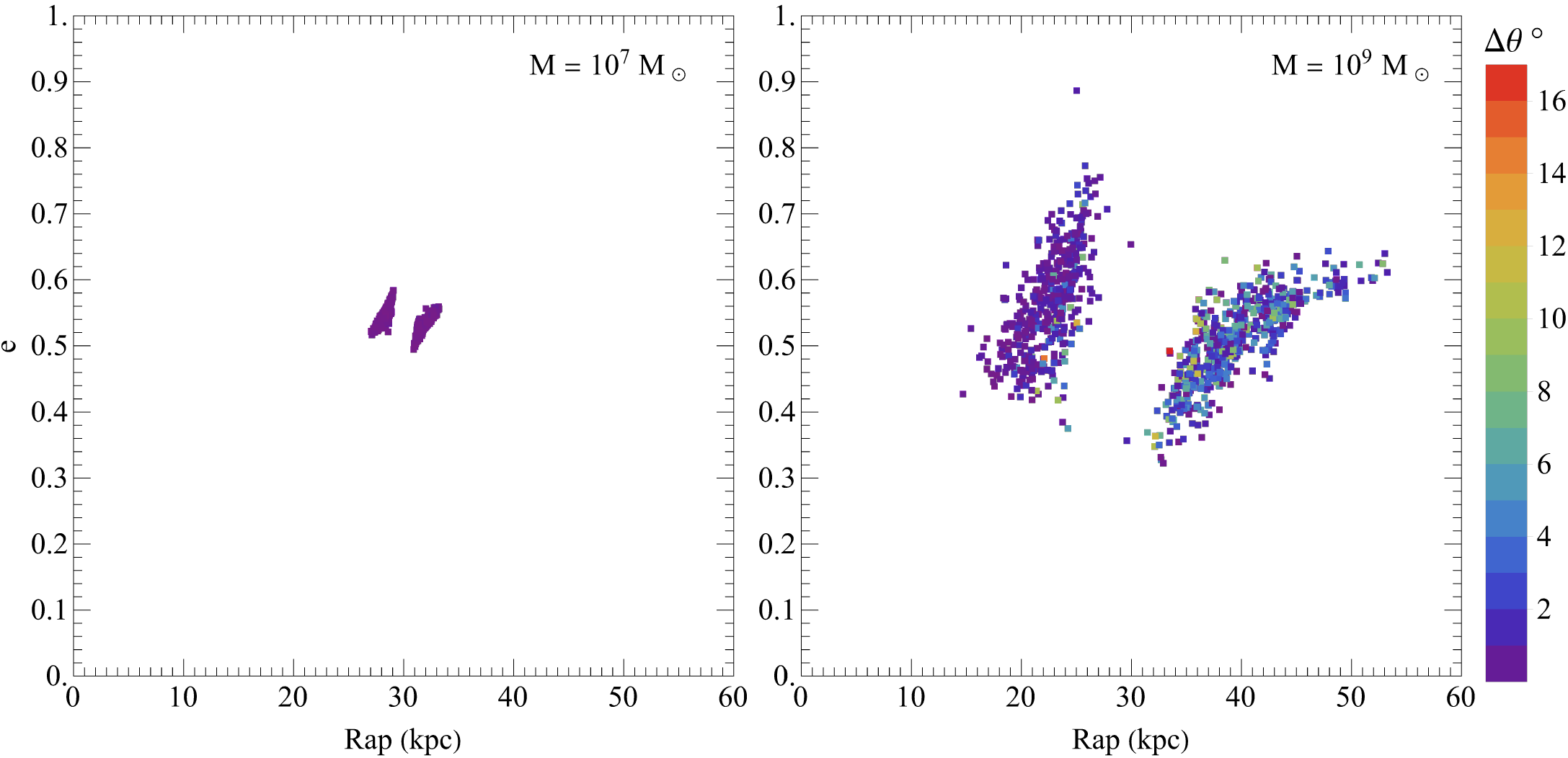}  	
	\caption{Orbital parameters of 1000 GCs escaped from two DSGs located on the MW disc after $8\Gyrs$ in the H1 model. The eccentricities of DSGs are 0.5. The colour coding represents the orbital inclination of GCs with respect to the DSG.
 The left and right panels correspond to DSG masses of $10^7$ and $\Msun[9]$, respectively.} 
	\label{fig:OrbitalElements_Mass}
\end{figure*}

We consider the MW, the DSG, and the GC as point masses whose masses are $M$, $m_1$, and $m_2$, respectively. For simplicity, we restrict their motion to two dimensions only, i.e. the $X-Y$ plane. Initially, the GC is bound to the DSG and forms a binary with a semi-major axis of $a_{12}$ and an eccentricity of $e_{12}$. In addition, the common centre of mass (CM) of the DSG-GC system orbits the MW, with an eccentricity and semi-major axis of $\eCM$ and $\aCM$, respectively. Once the GC becomes unbound, the DSG and the GC have their own orbits around the MW with orbital parameters of $(a_{1},e_{1})$ and $(a_{2},e_{2})$, respectively. The conservation of energy yields
\begin{equation} \label{eq:conser_energy}
     E_{(\mathrm{DSG-GC})} + E_{(\mathrm{MW-CM})} = E_{(\mathrm{MW-DSG})} + E_{(\mathrm{MW-GC})}
\end{equation}
where
\begin{multline}
  E_{(\mathrm{MW-CM})} = -\frac{G \left(m_1+m_2\right) M}{2 \aCM} \\
  E_{(\mathrm{DSG-GC})} = -\frac{G m_1 m_2}{2 a_{12}} \\ 
  E_{(\mathrm{MW-DSG})} = -\frac{G m_1 M}{2 a_1} \\
  E_{(\mathrm{MW-GC})} = -\frac{G m_2 M}{2 a_2} \\
\end{multline} 
Since the motion of objects is on the $X-Y$ plane only, the angular momentum is in the $Z$ direction ($\vec{L}=L\hat{z}$). The angular momentum for a two-body system is
\begin{equation} \label{eq:angular-momentum}
 L = \mu  \sqrt{a \left(1-e^2\right) G M}
\end{equation} 
where $\mu$ and $M$ are the reduced mass and the total mass, respectively. The conservation of angular momentum yields
\begin{equation} \label{eq:conser_angular-momentum}
     L_{(\mathrm{DSG-GC})} + L_{(\mathrm{MW-CM})} = L_{(\mathrm{MW-DSG})} + L_{(\mathrm{MW-GC})} \\
\end{equation}
where
\begin{multline}
 L_{(\mathrm{MW-CM})} = \frac{(m_1+m_2)M}{m_1+m_2+M}\sqrt{G \aCM \left(1-\eCM^2\right) \left(m_1+m_2+M\right)} \\
 L_{(\mathrm{DSG-GC})} = \frac{m_1 m_2}{m_1+m_2}\sqrt{a_{12} \left(1-e_{12}^2\right) G \left(m_1+m_2\right)} \\ 
 L_{(\mathrm{MW-DSG})} = \frac{m_1 M}{m_1+M}\sqrt{a_1 \left(1-e_1^2\right) G \left(m_1+M\right)} \\
 L_{(\mathrm{MW-GC})} = \frac{m_2 M}{m_2+M}\sqrt{a_2 \left(1-e_2^2\right) G \left(m_2+M\right)} \\
\end{multline}

\par Given the values for $\{M, m_1, m_2, \aCM, a_{12}, \eCM, e_{12}, e_1, e_2\}$ one can solve the system of equations \ref{eq:conser_energy} to \ref{eq:conser_angular-momentum} for unknowns $a_1$ and $a_2$. As an example, suppose $M=1.6\times\Msun[12]$ (H1 model), $m_1=\Msun[8]$, and $m_2=\Msun[5]$. The GC initially orbits the DSG with $(a_{12}=1\kpc,e_{12}=0)$. Moreover, the DSG-GC system initially orbits the MW on an orbit with $(\aCM=50\kpc,\eCM=0)$. Owing to the negligible mass of the GC compared to the mass of DSG, it has a marginal effect on the DSG orbit upon becoming unbound, hence $e_1\approx\eCM$. We assume that the orbit of the GC after the escape is circular ($e_2=0$). Solving the system of equations, yields two solutions for unknowns $a_1$ and $a_2$. One solution is $\{a_1=49.9958, a_2=54.3845\}$, describing an orbit on which the cluster reaches higher distances. The other solution is $\{a_1=50.0042, a_2=46.0204\}$ corresponds to an orbit with a smaller semi-major axis for the GC. This is a pattern that we observe in all of our simulations, i.e. there exist two populations of runaway GCs, scattered either closer to or further from the MW, with respect to their host DSG. This dichotomous pattern is due to GCs escaping in the vicinity of the two Lagrange points ${\mathrm L}_1$ and ${\mathrm L}_2$.   

\par It is worthwhile to mention that for some given values, these equations do not have any solutions. For example, consider the case of $\eCM=0.5$ and $e_{2}=0$ for which the equations are unsolvable. In \secref{sec:eff_mass}, we will examine the effect of $m_1$ and $m_2$ on the distribution of GCs.

\subsubsection{The effect of DSG mass on the distribution of orbital parameters of runaway GCs}\label{sec:eff_mass}

To illustrate the effect of the DSG mass on the distribution of orbital parameters of runaway GCs, we assume that the values of $M$, $m_2$, $a_\text{CM}$, $a_{12}$, and $e_{12}$ are given and equal to those of \secref{sec:dichotomy}. If we consider two values for $e_\text{CM}\in\{0.0, 0.5\}$, then for different masses of a DSG with different eccentricities of GCs, we obtain two values for $a_2$. \figref{fig:e_vs_a} shows the values of $a_2$ for each $e_2$ in the range of $10^8$ to $\Msun[10]$ for $m_1$. As $m_1$ increases, the dispersion in the semi-major axis ($a_2$) of GCs around the MW increases. As an example, for $e_\text{CM}=0.0$ and $e_2=0.3$, if $m_1=\Msun[8]$, then $a_2\in\{36, 75\}\kpc$, whereas for $m_1=\Msun[10]$ we have $a_2\in\{25, 114\}\kpc$. As another example, if $e_\text{CM}=0.5$ (the top panel of \figref{fig:e_vs_a}) and $m1<10^{10}\Msun$, the equations do not have any solutions for $a_2$ and $a_1$ if $e_2<0.5$. In all these calculations, we assume $m_2= \Msun[5]$. If we set $m_2=\Msun$, the values of $a_2$ do not change much. This means that the distribution of GCs that escaped from DSG is similar to the field stars that possibly escaped from it (cf. \citealt{Khalaj2016}).

\par Moreover, we consider two DSGs with masses of $10^7$ and $\Msun[9]$. The potential profile of these DSGs is the Plummer model, and we determine their scale lengths $(\alpha)$ such that the mass of the DSG divided by $\alpha^3$ is equal to $0.5\Msun\pc^{-3}$. This is a reasonable density for the MW DSGs, because according to Table 3 of \citetalias{Rostami2022}, the average total mass density of DSGs is close to $0.5\Msun\pc^{-3}$. We placed these DSGs at a distance of $30\kpc$ on the MW disc and selected their velocities such that their eccentricities are $e=0.5$. We use the H1 model for the simulation and obtain the orbits of 1000 runaway GCs from each of the DSGs after $8\Gyrs$. The orbital parameters of these GCs can be seen in \figref{fig:OrbitalElements_Mass}. It is evident that the dispersion of orbital parameters of runaway GCs increases as the mass of the DSG increases. In other words, for heavier DSGs, the orbits of escaped GCs have a lower similarity to their host DSG orbit. The dynamical masses of most MW DSGs are less than $\Msun[8]$. As a result, the orbits of the escaped GCs are expected to be very similar to the orbit of their host DSG. 

The Sgr is one of the DSGs that is close to the centre of the MW. Most of the other DSGs are located at larger radii, and consequently, the MW has stripped a smaller fraction of their initial mass in comparison. As a result, their present-day mass is not significantly different from their initial mass. This means that the choice of a static potential is a good approximation for simulating these DSGs. Moreover, one can argue that dynamical friction has a marginal effect on these DSGs due to their low mass, compared to e.g. the Sgr or the LMC. We showed that there is a positive correlation between the dispersion of the orbital parameters of the runaway GCs, and the mass of their host DSGs. Combined with the fact that the Sgr is heavier than other DSGs (except for the LMC and the SMC), we can conclude that the Sgr filters in the static models can be considered as a high limit for the filters of other DSGs. This justifies our methodology in \secref{sec:prob_DSG}. It should be noted that repeating this test for the lower-mass DSGs leads to much narrower filters. Future investigations based on more accurate data of density profiles and orbital parameters of other 38 MW DSGs can improve the results.

\begin{figure}
  \centering
  \includegraphics[width=1\linewidth]{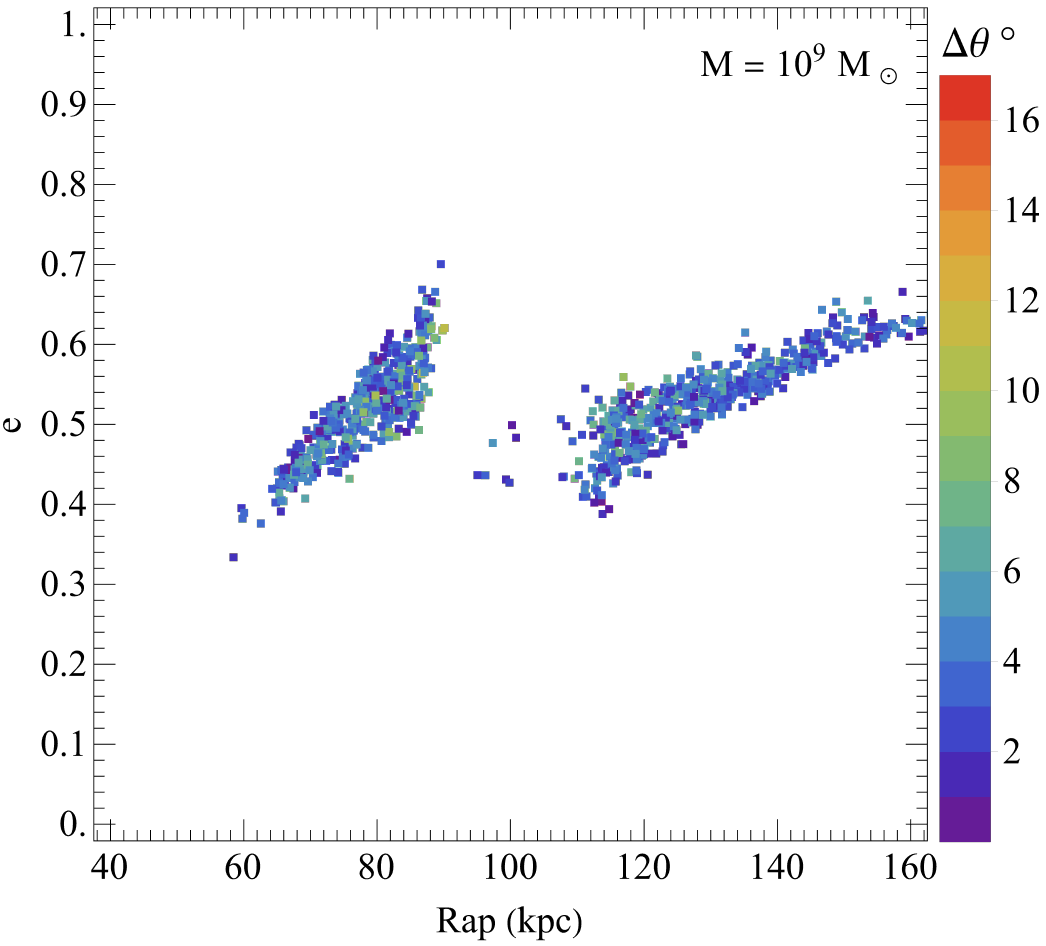}
  \caption{Same as the right panel of \figref{fig:OrbitalElements_Mass} but for DSG orbits at a higher apogalactic distance $\Rap=100\kpc$.}
  \label{fig:OrbitalElements_Rap}
\end{figure}

\begin{figure*}
    \centering
    \includegraphics[width=1\linewidth]{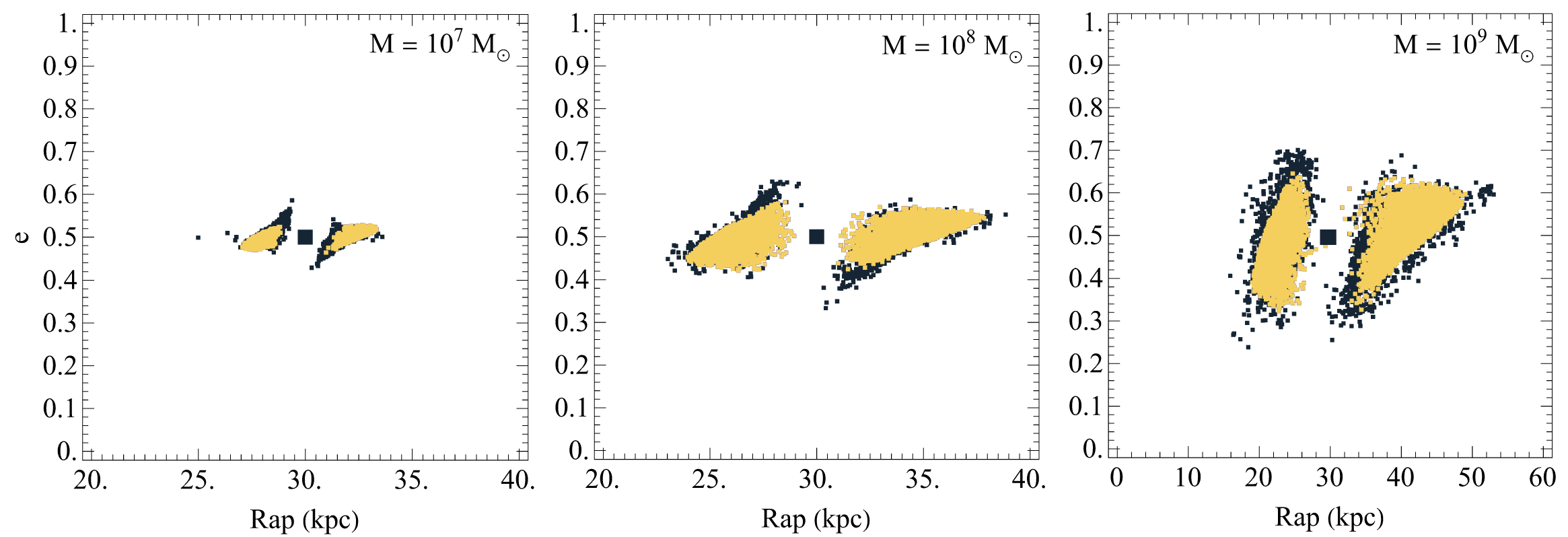}  	
	\caption{Distribution of $\Rap$ and $e$ of runaway particles from three DSGs located on the MW disc with orbital parameters of $\Rap=30\kpc$ and $e=0.5$, obtained from \textsc{NBODY6} (yellow) and our three-body method (black), for three DSG masses of $10^7$, $10^8$ and $\Msun[9]$.}
	\label{fig:comp-Nbody}
\end{figure*}

\subsubsection{The effect of DSG orbital distance on the distribution of orbital parameters of runaway GCs}\label{sec:eff_ra}

We assume a DSG with the mass of $\Msun[9]$, an orbital eccentricity of $e=0.5$, and two different apogalactic distances of $R_\mathrm{ap}=$ 30 and 100 $\kpc$. Similar to \secref{sec:eff_mass}, we use the H1 model and obtain the orbits of 1000 runaway GCs after $8\Gyrs$. \figref{fig:OrbitalElements_Rap} shows the orbital parameters of runaway GCs. A comparison with the right panel of \figref{fig:OrbitalElements_Mass}, shows that as $\RapDSG$ increases the values of $\FDSGE$ and $\FDSGTheta$ do not change. In the case of $\Rap=30$ kpc, $\FDSGR$ is $\sim20\kpc$, while for $\Rap=100$ $\FDSGR$ increases to approximately $60\kpc$. This indicates that the values of $\RapDSG$ and $\FDSGR$ are positively correlated.

\subsection{Comparison of direct \textit{N}-body and our large ensemble of three-body simulations: distribution of tail stream of DSGs\label{sec:Nbody-N3body}}

First, we use \textsc{NBODY6} \citep{Aarseth2003} direct \Nbody code, to obtain orbital parameters of runaway stars of DSGs due to the Galactic tidal field, assuming three DSG masses of $10^7$, $10^8$, and $\Msun[9]$. All modelled DSGs embody $N=10^5 $ equal-mass particles. The initial positions and velocities of the particles in the DSGs are set such that the mass-density obeys a Plummer profile in virial equilibrium. We determine the corresponding scale lengths $(\alpha)$ such that the mass of the DSG divided by $\alpha^3$ is equal to $0.5\Msun\pc^{-3}$. The particles do not undergo stellar evolution. The DSGs move on an orbit with $e=0.5$ through a host galaxy that consists of three components, a central point-mass with $m=1.5\times\Msun[10]$, a \citet{Miyamoto1975} disc potential with numerical constants $\Md = 5\times\Msun[10]$, $a = 4 \kpc$ and $b = 0.5 \kpc$ (see \equref{eq:miyamoto}), and a logarithmic potential for the halo as follows
\begin{equation}\label{eq:phi_halo}
    \phih  = {\frac{1}{2} \Vinfty^{2} \ln\left(\Rc^2 + R^2\right)},  	
\end{equation}
The constant $\Rc$ is chosen such that the combined potential of the three components yields a circular velocity of $\Vinfty=220\kms$ in the disc plane at a distance of $8.5\kpc$ from the galactic centre. We placed these DSGs at a distance of $30\kpc$ on the MW disc. At the end of simulations, we obtain the orbital parameters of runaway particles, i.e. particles of which final distances exceed $2\times\Rtidal$ from the centre of their host DSG.

Next, using our three-body method we perform a simulation for the H1 model, in which the MW potential is adjusted to match that of \textsc{NBODY6} and obtain the orbital parameters of 2500 runaway particles from the DSGs.

\figref{fig:comp-Nbody} compares $\Rap$ and $e$ of runaway particles from modelled DSGs as determined by \textsc{NBODY6} and our three-body methods. Evidently, the resultant distributions from different methods perfectly match. In particular, the region that runaway particles occupy in the orbital parameter space expands as the DSG mass increases. The observed consistency between the two approaches extends beyond $\Rap$ and $e$, and includes the inclination of runaway particles as well. As a result, our large ensemble of three-body methods are an acceptable alternative for direct \Nbody methods which are computationally expensive, and therefore can be readily utilized to obtain the distribution of tail streams of DSGs.

\section{Results}\label{sec:results}
\subsection{Runaway GCs of the Sgr}\label{sec:filter_sgr}

\begin{figure*}
    \centering
    \includegraphics[height=0.9\textheight]{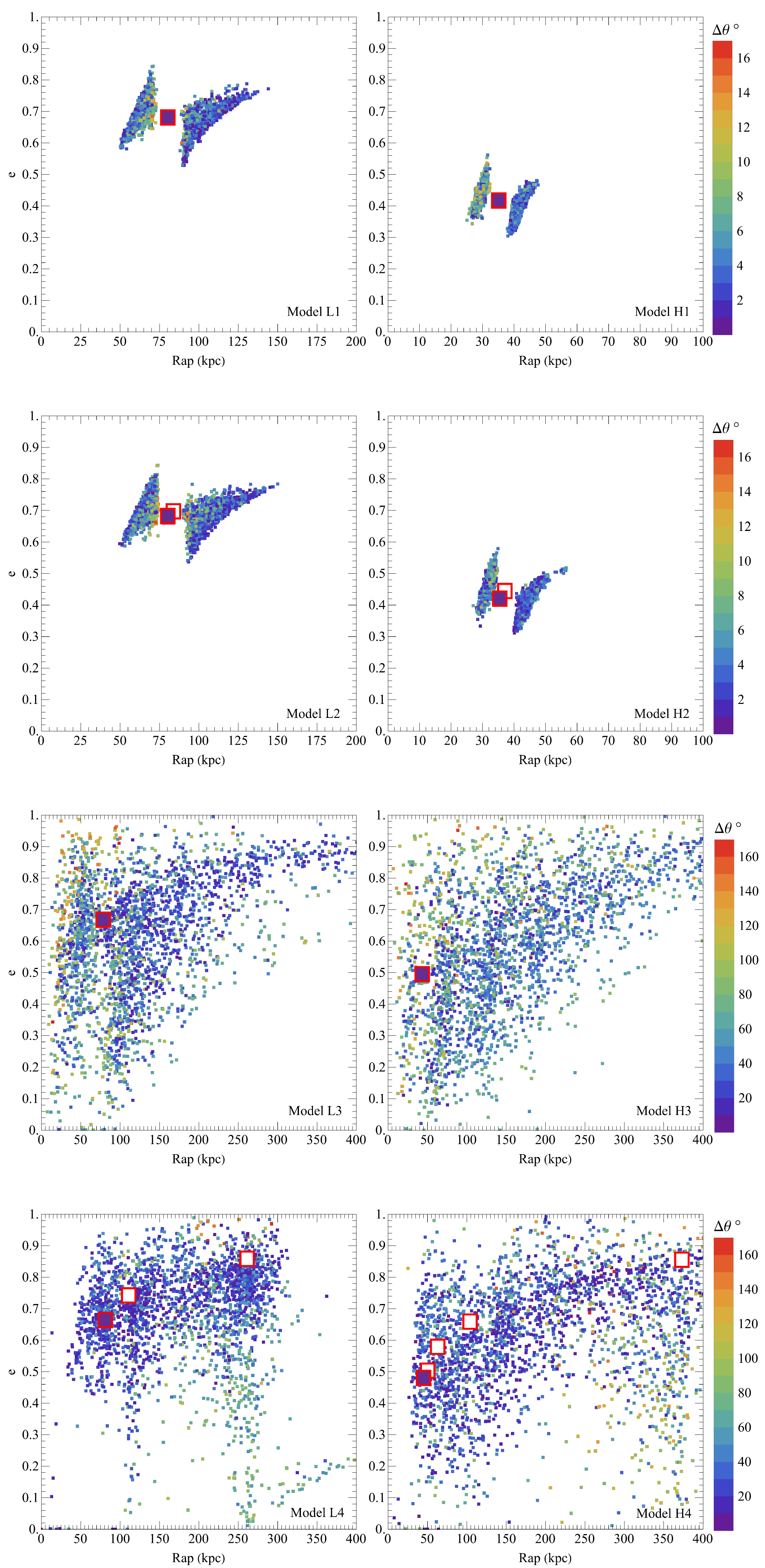}  	
    \caption{The orbital parameters of 2500 runaway GCs, from the Sgr in eight simulation models. The horizontal axis is the apogalactic distance ($\Rap$) in $\kpc$ and the vertical axis is the orbital eccentricity of these GCs. The colour coding corresponds to the orbital inclination of the GCs in degrees, with respect to the orbital plane of the Sgr. The large squares mark the Sgr. For models with dynamical friction, the position of the Sgr in the parameter space changes as a function of time. This has been shown by the magenta-filled square for the present day, and the white-filled squares for earlier epochs. For H2, H4, L2, and L4 models, the orbital parameters of the Sgr change over time as a result of dynamical friction.} 
    \label{fig:OrbitalElements_Sgr}
\end{figure*}

\begin{figure}
  \centering
  \includegraphics[width=0.9\linewidth]{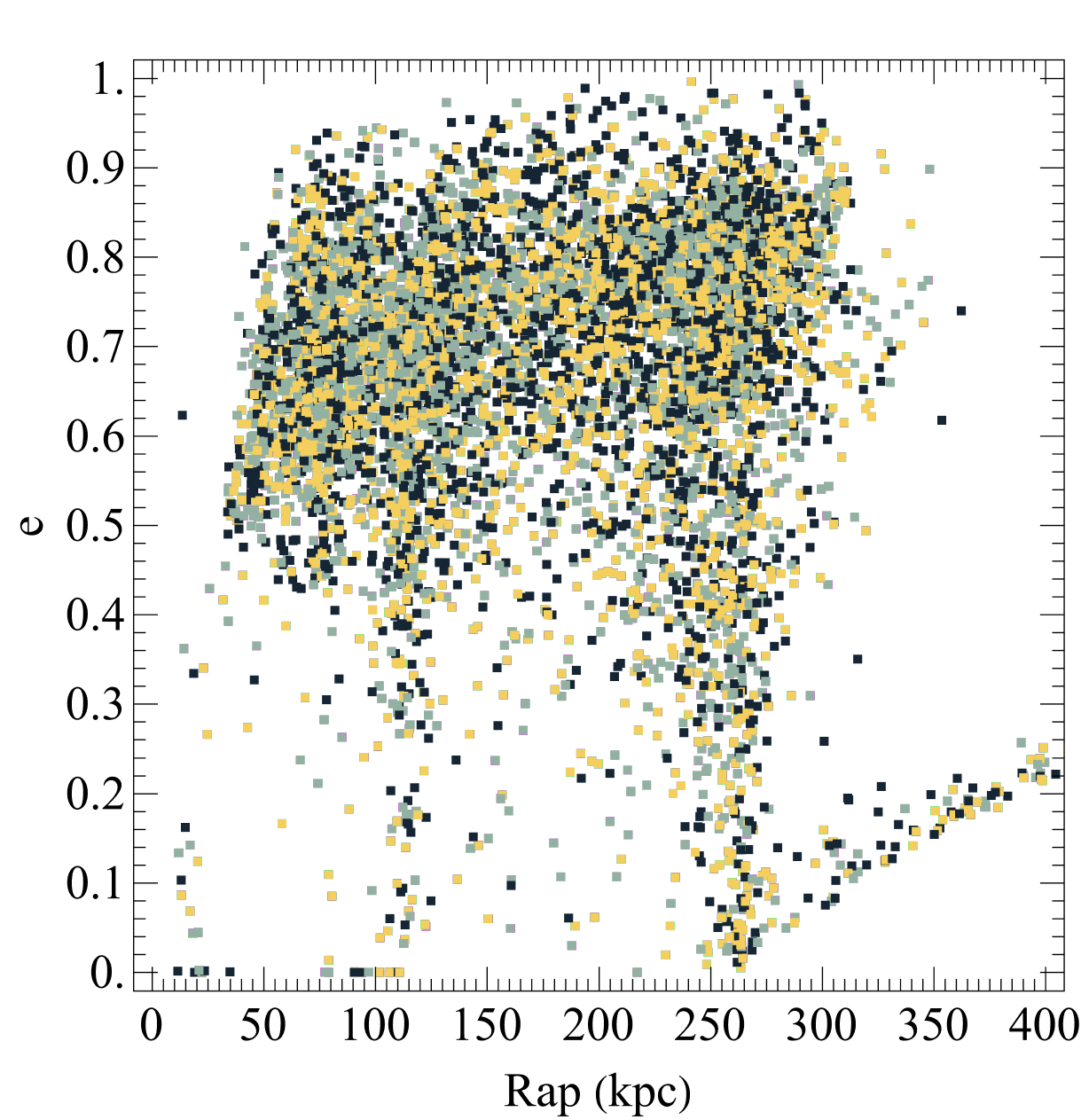}
  \caption{The present-day distribution of orbital parameters of runaway GCs for the L4 model assuming three different values for the ratio of baryonic to dark-halo scale lengths, for the initial Plummer distribution of GCs, i.e. $r_\mathrm{sc,bar}=\kappa r_\mathrm{sc,halo}$, where $\kappa=1.0$ (black), $\kappa=0.5$ (green), and $\kappa=0.25$ (yellow). The distribution of orbital parameters of runaway GCs is not sensitive to the adopted initial GCs spatial distribution.}
  \label{fig:GCs-dis}
\end{figure}

\figref{fig:OrbitalElements_Sgr} shows the orbital parameters of 2500 runaway GCs in each of the adopted simulation models. It evidently depicts two populations of GCs, one with $\Rap\leq\Rap(\mathrm{Sgr})$ and another one with $\Rap>\Rap(\mathrm{Sgr})$. This has been already shown semi-analytically in \secref{sec:analytical_met}. This dichotomy resembles a butterfly pattern in which each wing corresponds to one of the aforementioned populations. Interestingly, one can see that the runaway GCs occupy a wider range in the dynamic models (the bottom two rows), compared to static ones (the top two rows). This is due to the larger initial mass of the Sgr in dynamic models, in which some of the runaway GCs can travel as far as $400\kpc$ away from the MW, while some others can reach the MW bulge. Moreover, the orbital inclination can be as large as $170\degr$. This implies a flip in the angular momentum vector, meaning that some runaway GCs have retrograde orbits around the MW. This can be compared with the static models (L1, L2, H1, H2), where the maximum value of the orbital inclination is about $17\degr$.   

\par In the H4 and L4 models, where dynamical friction is also at play, the Sgr starts with a larger value of $\Rap$ and $e$. It then spirals inward and drags (runaway) GCs along with itself. This can be seen in the bottom row of \figref{fig:OrbitalElements_Sgr}, where the distribution of runaway GCs follows the Sgr trajectory. As expected, the present-day distribution of orbital parameters of runaway GCs obtained from the static model (L2) over $8\Gyr$ yields similar results to that of the dynamic model (L4) over the last $1\Gyr$. We also investigated the possible interference of the LMC on the orbital parameters of the Sgr runaway GCs. We considered masses of $10^{10}$ to $\Msun[11]$ \citep{Erkal2019} with and without dynamical friction for the LMC. The distribution of the Sgr runaway GCs remained almost unchanged in all these cases due to the large distance of the Sgr from the LMC.

To calculate the equations of motion for GCs, we need to consider the total mass (stellar mass of the galaxy + dark-halo mass) of DSGs. The GCs follow the baryonic component of the DSGs, i.e. the GC system size is close to the DSG size (e.g. \citealt{Caso2019}). Estimating the initial size of the stellar component of a DSG is not trivial. As a result, we consider three different radii for the distribution of GCs to investigate the sensitivity of the results to this choice. \figref{fig:GCs-dis} shows the present-day distribution of orbital parameters of runaway GCs for L4 model assuming three different values for the ratio of barynoic to dark-halo scale lengths, for the initial Plummer distributions of GCs, i.e. $r_\mathrm{sc,bar}=\kappa r_\mathrm{sc,halo}$, where $\kappa\in\{0.25, 0.5, 1.0\}$. Owing to the fact that GCs located at large radii from the centre of their host DSG are only loosely bound, one might expect a larger dispersion of orbital parameters in the parameter space for larger $\kappa$ values. Interestingly, this is not what we observe in our simulations (\figref{fig:GCs-dis}). Despite the fact that the escape rate of GCs are significantly different in the three $\kappa$ cases, their distribution in $R_\mathrm{ap}$ and $e$ is similar (\figref{fig:GCs-dis}). This is due to the initial bounding condition of GCs (\secref{sec:methodology}). In particular, for $\kappa=1.0$ a large fraction of GCs are initially located in the outer part of the DSG. Such GCs do not satisfy the initial bounding criterion, i.e. they do not remain within $2\times\Rtidal$ for about four times their orbital period. Therefore, they are not counted as runaway GCs and excluded. In other words, GCs that ultimately contribute to the distribution of runaway GCs, form a sub-system with an effective $\kappa$ value which is less than 1.0. On the other hand, for $\kappa=0.25$, the GCs mainly reside in the inner region of the DSG, which has a high escape speed. Therefore, runaway GCs emanating from these areas must have a high speed. This leads to a more scattered distribution of orbital parameter for such DSGs, i.e. such DSGs have a larger effective $\kappa$ value. Therefore, cases of $\kappa=1.0$ and $\kappa=0.25$ yield distributions which are only marginally different from $\kappa=0.5$, if any at all. As a result, the distribution of orbital parameters of runaway GCs does not depend on the initial distribution of GCs and the most important factor in the dispersion of runaway GCs is the total mass of the DSG. In addition, in \secref{sec:GC_asso_LMC} we demonstrate that it does not even depend on the adopted density profile. Hereafter, we assume $\kappa=0.5$ for all simulations.

\subsection{GCs associated with the Sgr}\label{sec:GC_asso_sgr}
\begin{table}
	\centering
	\begin{tabular}{ccccccccc}
	\hline
        & & & & & & \\
        GC & H1 & H2 & H3 & H4 & L4 & L3 & L2 & L1\\ 
        & & & & & & \\
	\hline
                  
         NGC 6715 (M54) & 1 & 1 & 1 & 1 & 1 & 1 & 1 & 1\\         
         Ter 7 & 1 & 1 & 1 & 1 & 1 & 1 & 1 & 1\\         
         Arp 2 & 1 & 1 & 1 & 1 & 1 & 1 & 1 & 1\\         
         Ter 8 & 1 & 1 & 1 & 1 & 1 & 1 & 1 & 1\\ 
         Whiting 1 & 2 & 1 & 1 & 1 & 1 & 1 & 2 & 2\\
         Pal 12 & 2 & 2 & 1 & 1 & 1 & 1 & 2 & 2\\		
         NGC 2419 & - & - & 1 & 1 & 1 & 1 & 2 & 2\\		
         NGC 5466 & - & - & 1 & 1 & 1 & 2 & 2 & 2\\ 
         NGC 5634 & - & - & 2 & 1 & 2 & 1 & - & -\\
         NGC 4590 & - & - & 1 & 2 & 2 & 1 & - & -\\
         Rup 106 & - & - & 2 & 2 & 2 & 1 & - & -\\         
         Pal 4 & - & - & 1 & 2 & 2 & 2 & - & -\\
         NGC 6101 & - & - & 1 & - & 2 & 2 & - & -\\
         NGC 5897 & - & - & 2 & 2 & - & 2 & - & -\\
         NGC 6235 & - & - & 2 & 2 & - & 2 & - & -\\
         NGC 6934 & - & - & - & - & 2 & 2 & - & -\\         
         NGC 6426 & - & - & 2 & 2 & - & - & - & -\\
         IC 4499 & - & - & 2 & - & - & 2 & - & -\\
	\hline		
	\end{tabular}
	\caption{List of GCs that could be associated with the Sgr in all simulation models. The numbers indicate the association flags. Flag 1 GCs have the highest probability of being associated with the Sgr (ref. \secref{sec:sgr-assoc}).}
	\label{tab:Sgr_GCs}
\end{table}

\begin{figure*}
	\centering
	\includegraphics[width=1\linewidth]{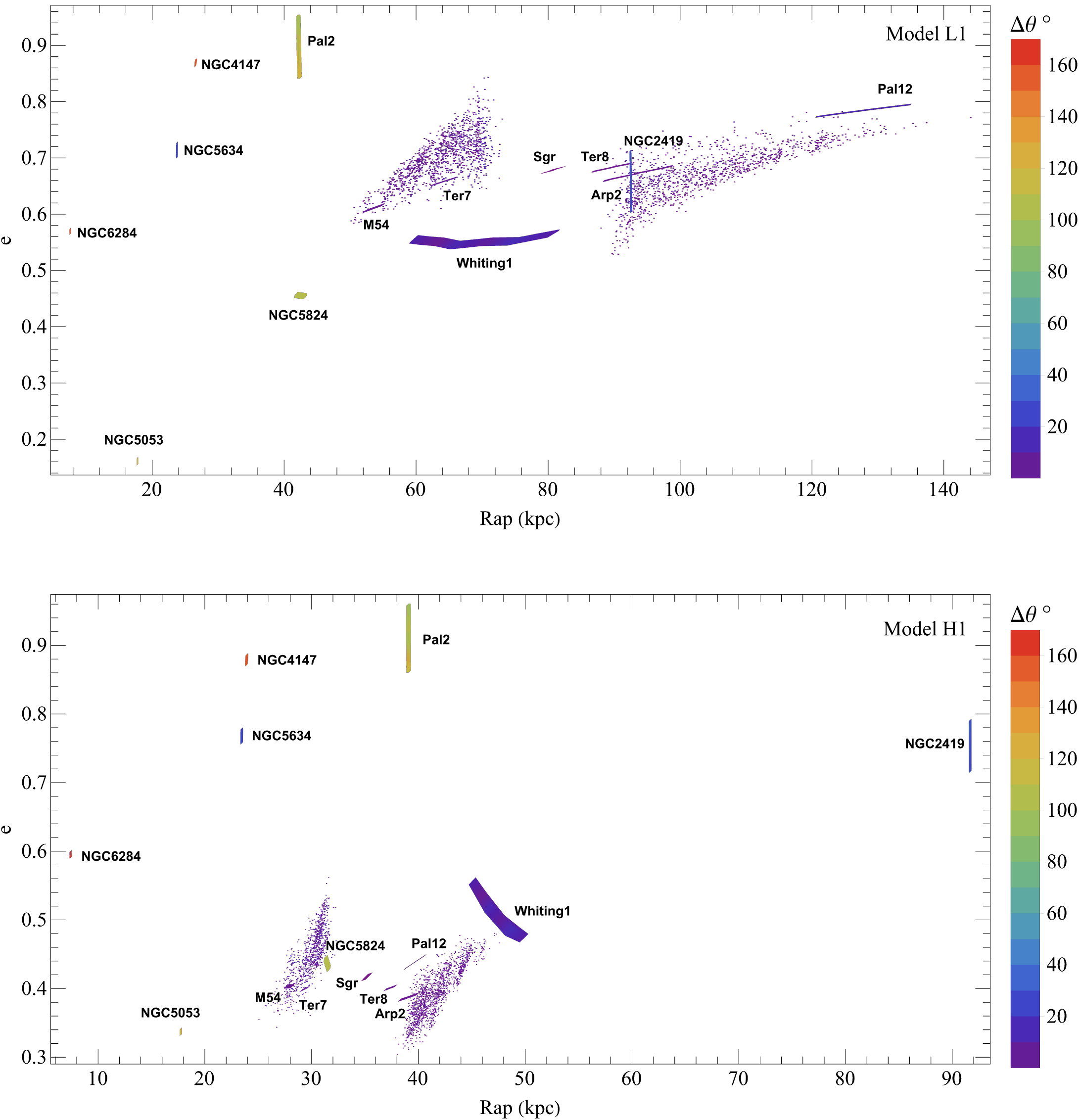}  	
	\caption{Orbital parameters of the Sgr runaway GCs (dots) and 13 MW GCs in L1 and H1 models, which have been previously suggested to be associated with the Sgr. The area occupied by each MW GC is due to its uncertainty of proper motions and the line-of-sight velocity. The Sgr is also depicted, considering the uncertainty of its observational data.}  
	\label{fig:OrbitalElements_MW}
\end{figure*}

Using the runaway GCs of the Sgr given in \secref{sec:filter_sgr} and the method described in \secref{sec:prob_sgr}, we determine the probability of association between MW GCs and the Sgr. \figref{fig:OrbitalElements_MW} shows the orbital parameters of the Sgr runaway GCs in H1 and L1 models, as well as those of 13 MW GCs, studied in \citet{massari2019}, \citet{belazini2020}, and \citet{LAW2010b}, considering their uncertainties. These MW GCs are believed to have escaped from the Sgr. The Sgr orbital uncertainty is very small, hence it occupies a small area in the parameter space. As evident from the figure, many of the MW GCs that are thought to have originated from the Sgr, are far away from the Sgr runaway GCs, in terms of the orbital parameters. In particular, NGC 6284, NGC 5053, NGC 4147, NGC 5824, and Pal 2 are highly unlikely to have originated from the Sgr.  

\par For the H1 model, only six GCs have similar orbits to the Sgr runaway GCs. These are Whiting 1, NGC 6715 (M54), Ter 7, Arp 2, Ter 8, and Pal 12. 
NGC 5824 lies within the range of Sgr filters in terms of $e$ and $R_\text{ap}$. However, its orbital inclination is $\sim100\degr$, whereas the maximum orbital inclination of the Sgr runaway GCs in the H1 is only $17\degr$. In the L1 model, in addition to the GCs mentioned for the H1, NGC 2419 and NGC 5466 are candidates for being associated with the Sgr. \tabref{tab:Sgr_GCs} lists the possible association of MW GCs with the Sgr in different models. In this list, we are reporting several Sgr-GCs associations for the first time. The dynamic models yield more associations as a result of their wider filter distributions.

\par In agreement with other studies, Whiting 1, NGC 6715 (M54), Ter 7, Arp 2, Ter 8, Pal 12, and NGC 2419 are likely to be associated with the Sgr. Moreover, our current study indicates that NGC 5466 has the possibility of an association in all L models as well as H3 and H4 models. In addition, Rup 106, NGC 4590, NGC 5634, and Pal 4 are likely to be associated with the Sgr in all dynamic models. NGC 6101, NGC 5897, NGC 6235, NGC 6934, NGC 6426, and IC 4499 are other GCs likely to be associated with the Sgr, albeit with a lower probability. Pal 4 is one of the GCs that has large orbital energy with an eccentricity of $e\approx0.9$, and our results indicate that it belongs to the Flag 2 category. This is in agreement with \figref{fig:OrbitalElements_Sgr}, where one expects GCs with $\Rap\approx300\kpc$ and an $e\approx0.9$ to have originated from the Sgr.

\par Assuming that M 54, Arp 2, Ter 7, and Ter 8 are still bound to the Sgr \citep{belazini2020}, \citetalias{Rostami2022} predicted that about 14 GCs could have escaped from the Sgr. Here we have introduced 18 GCs that are likely to be associated with the Sgr.

It should be noted that to calculate the orbit of the MW GCs, we have considered only the gravitational potential of the Galaxy and ignored the potential of other DSGs. As also shown in \figref{fig:OrbitalElements_MW}, this explains the reason why the orbital parameters of M~54, Arp~2, Ter~7, and Ter~8 lie within the range of orbital parameters of runaway GCs but they are not inside the Sgr. If we also consider the potential of Sgr for a cluster such as M~54, its location in \figref{fig:OrbitalElements_MW} will remain close to the Sgr after the evolution. This is in agreement with the fact that M~54 is thought to be the nuclear cluster of Sgr \citep{Ibata1994,Bellazzini2008}.

As a final remark, our simulations predict a concentration of runaway GCs for the Sgr at large distances of $\Rap\approx275\kpc$ for the L4 model and $\Rap\approx375\kpc$ for the H4 model. Both of these concentrations have high eccentricities of $e\approx0.8$ and relative inclinations of $\Delta\theta\approx20\degr$. These are not associated with any observed MW GCs which is an interesting discovery. We further elaborate upon this in \secref{sec:conclusion}.

\begin{figure*}
	\centering
	\includegraphics[width=1\linewidth]{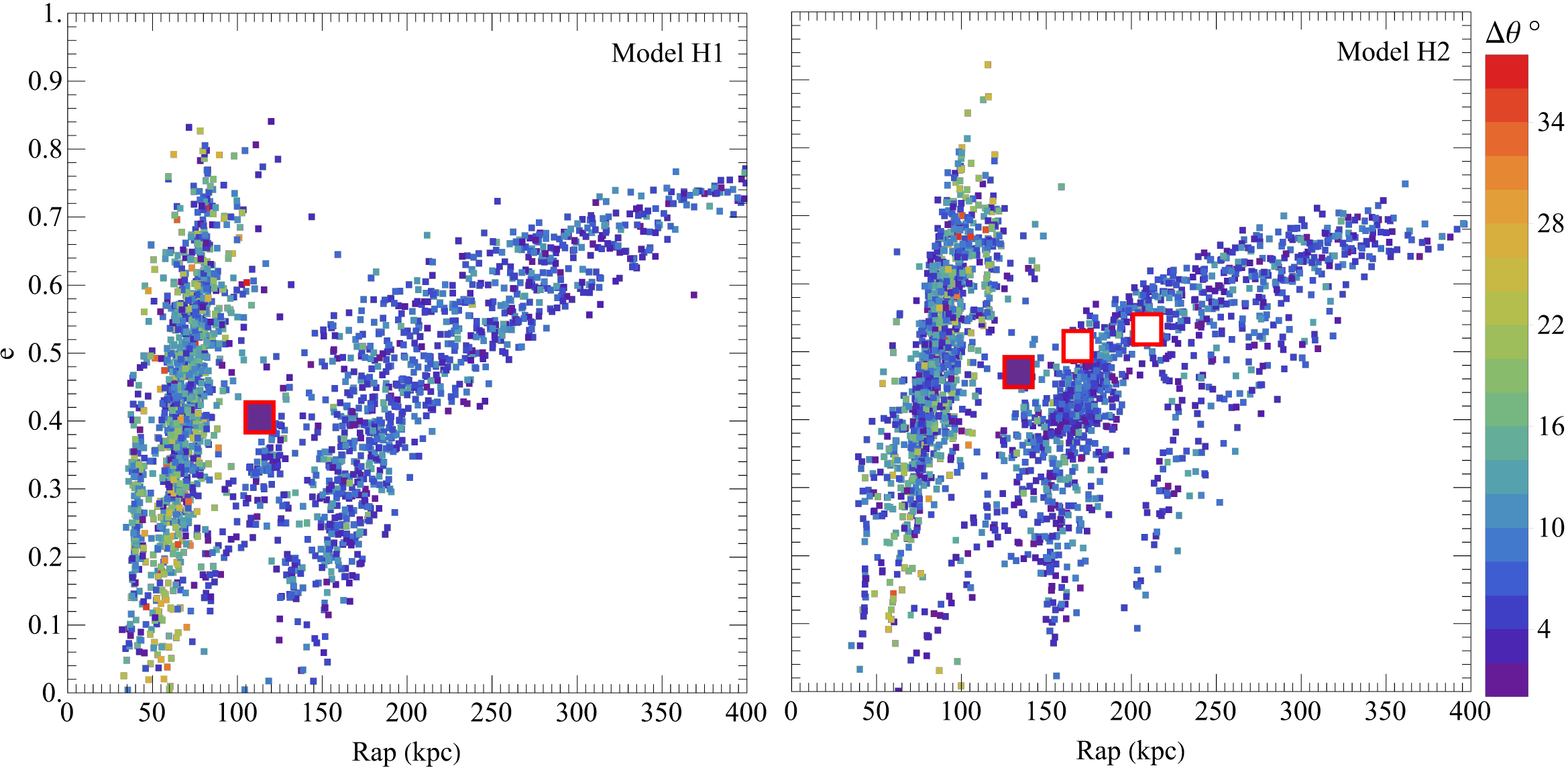}  	
	\caption{Same as \figref{fig:OrbitalElements_Sgr} but for the LMC runaway GCs in H1 (without the dynamical friction) and H2 (with the dynamical friction) simulation models. The large squares mark the LMC, and $\Delta\theta\degr$ has been measured with respect to the orbital plane of the LMC. Dynamical friction in H2 model, changes the orbital parameters of the LMC (shown as three squares). As a result, the distribution of the orbital parameters of GCs is more extended compared to H1 model. }  
	\label{fig:OrbitalElements_LMC}
\end{figure*}

\begin{figure}
  \centering
  \includegraphics[width=0.9\linewidth]{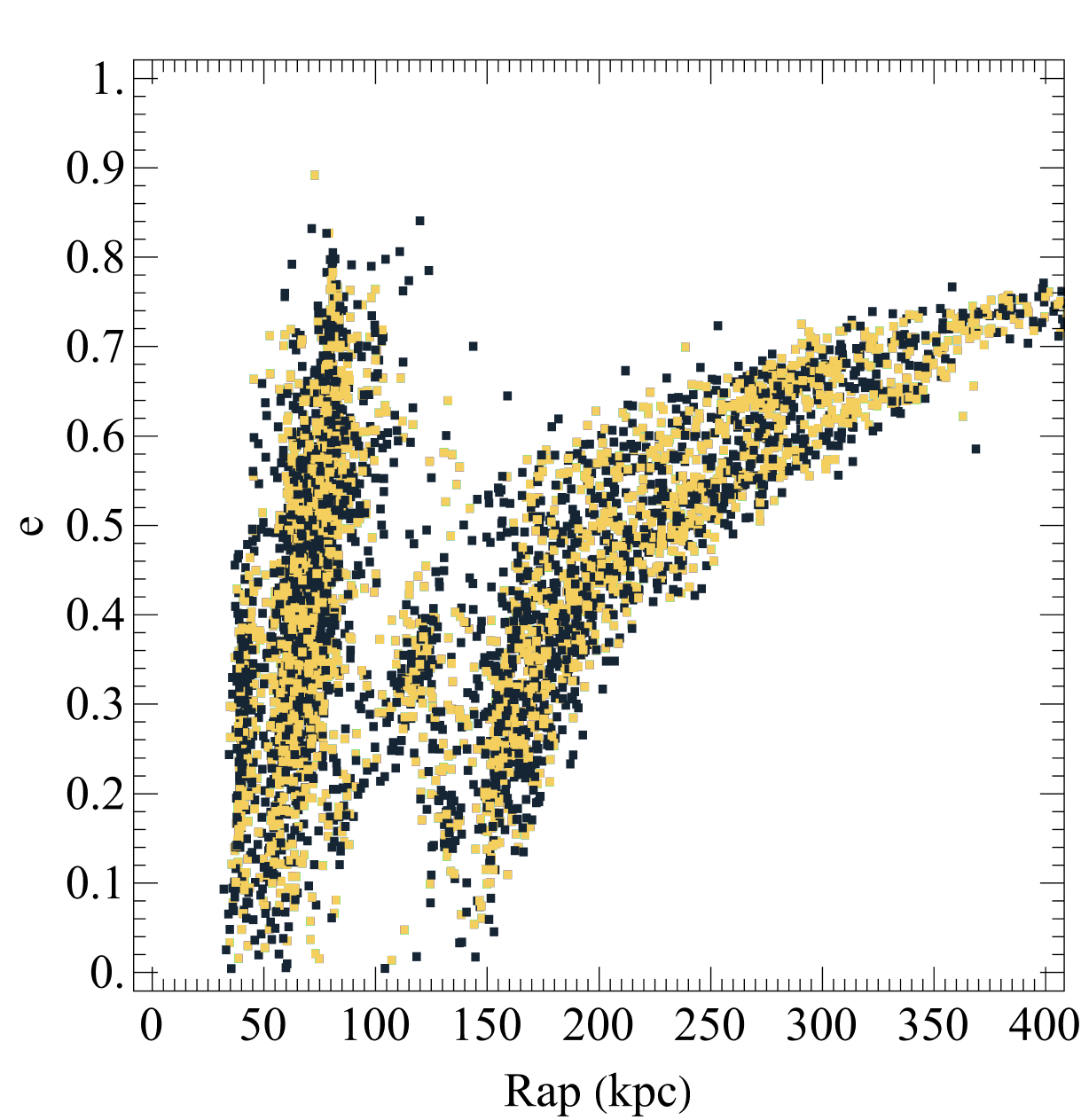}
  \caption{The orbital parameters of 2500 runaway GCs from the LMC in H1 model, assuming two different density profiles for the LMC, namely Plummer (black) and Hernquist (yellow).}
  \label{fig:Plum-Hern}
\end{figure}

\begin{figure}
  \centering
  \includegraphics[width=1\linewidth]{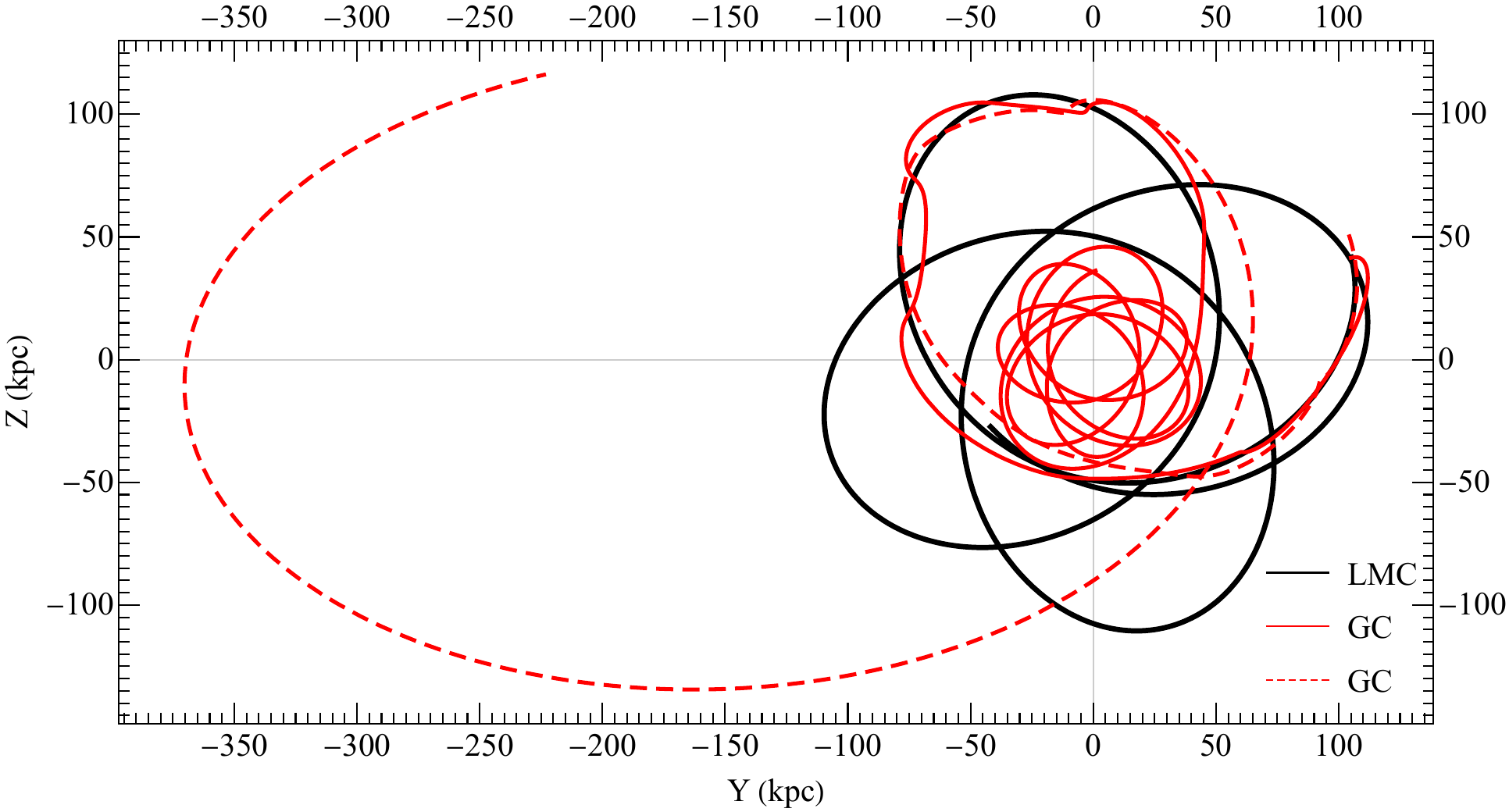}
  \caption{The orbit of the LMC (solid black line) as well as two runaway GCs after $8\Gyrs$ in the H1 model. The orbits are depicted for the $Y\text{-}Z$ plane of the Galactocentric coordinate. The solid red line shows an example where the GC is thrown towards the MW with $\Rap\approx30\kpc$. The dashed red line is an example where the $\Rap$ of the GC reaches $380\kpc$.}
  \label{fig:LMC_Orbit}
\end{figure}

\begin{figure*}
	\centering
	\includegraphics[width=1\linewidth]{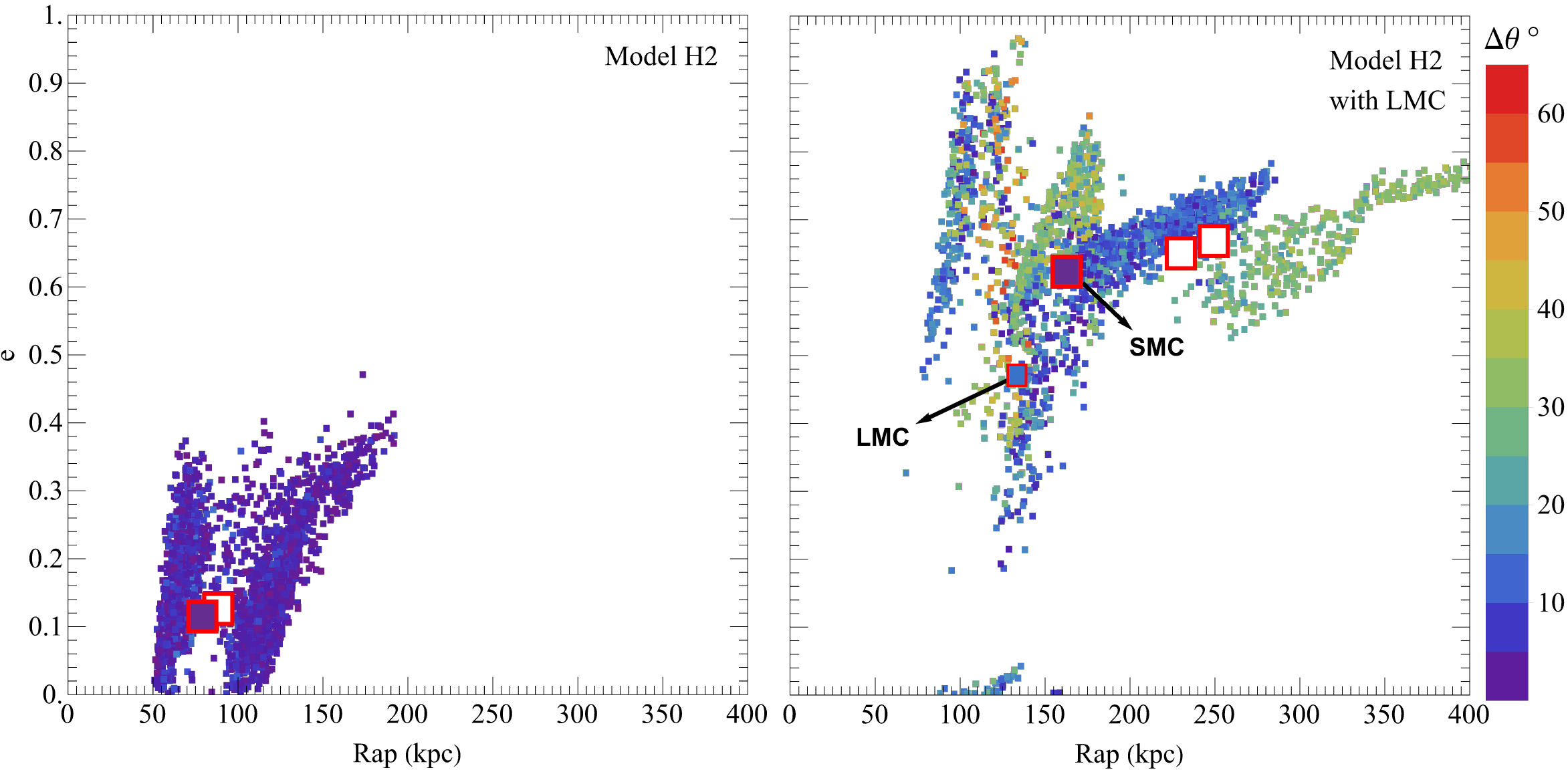}  	
	\caption{The orbital parameters of 2500 escaped GCs from the SMC in the H2 model without (left panel) and with (right panel) the presence of the LMC. The colour coding indicates the orbital inclination of the GCs with respect to the SMC orbit (in degrees). The orbital parameters of the SMC and the LMC are depicted by large squares. The white-filled squares show the orbital parameters of the SMC at earlier epochs as it changes due to the dynamical friction.}  
	\label{fig:OrbitalElements_SMC}
\end{figure*}

\begin{table*}
\begin{tabular}{cccccc|cccccc}
    \hline
        & & & & & & & & & & &\\
        DSG & GC  & H1 & H3 & L1 & L3 & DSG & GC & H1 & H3 & L1 & L3 \\
        & & & & & & & & & & &\\
	\hline 
	& & & & & & & & & & &\\
	\textbf{Antlia 2}    & NGC 4590	& - & - & 1 & 1 & \textbf{Fornax I}    & Crater   & 2 & 2 & 2 & 2 \\ 
	& & & & & & & & & & &\\
	
	\textbf{Boötes I}    & AM 4      & 2 & 2 & - & - & \textbf{Hercules I}  &	NGC 6426 & 2 & 2 & - & - \\	
	& & & & & & & & & & &\\
	
	                     & NGC 5824	 & 1 & 1 & 1 & 1 &                      & Eridanus   & - & - & 2 & - \\
	                     & & & & & & & & & & &\\
	\textbf{Boötes II}	 & Whiting 1 & 2 & - & - & - & \textbf{Horologium I}&  Pal 3     & 2 & 2 & 2 & 2 \\
	& & & & & & & & & & &\\
	                     &   Arp 2	 & 2 & - & - & - & \textbf{Hydrus I}    &  Pal 3     & 2 & 2 & - & - \\	
	                     & & & & & & & & & & &\\
		                 &  Pal 12	 & 2 & 2 & - & - &			            &  Pal 4     & 2 & 2 & - & - \\
		                 & & & & & & & & & & &\\
	\textbf{Carina I}    &	Pal 3	 & 1 & 2 & - & - & \textbf{Reticulum II}& NGC 5024	 & 1 & 1 & 1 & 1 \\
	& & & & & & & & & & &\\
	\textbf{Carina III}	 & Pyxis	 & 1 & 1 & 1 & 2 & \textbf{Segue 1}	    &NGC 6101	 & 1 & - & - & - \\	
	& & & & & & & & & & &\\
    \textbf{Crater II}	 & Pyxis	 & 2 & 2 & - & - & \textbf{Sextans I}	&NGC 2419	 & - & - & 2 & - \\
    & & & & & & & & & & &\\
	                     & NGC 4590	 & - & - & 2 & 2 & \textbf{Tucana II}	&Eridanus	 & - & 2 & 2 & 2 \\
	                     & & & & & & & & & & &\\
	                     & NGC 7492	 & - & - & 2 & 2 & \textbf{Triangulum II}&Eridanus	 & 2 & 2 & - & - \\	
	                     & & & & & & & & & & &\\
    \textbf{Draco I}	 & Pal 3     & 2 & 2 & 2 & 2 & \textbf{Tucana III}	 &NGC 5466	 & 1 & 1 & - & - \\
    & & & & & & & & & & &\\
	                     & Pal 4	 & 2 & 2 & 2 & 2 &                       &NGC 1261	 & 2 & 1 & 1 & 1 \\
	                     & & & & & & & & & & &\\
                	     & NGC 5024	 & - & - & 1 & 1 & \textbf{Ursa Major II}&Pal 13	 & - & - & 2 & 2 \\
                	     & & & & & & & & & & &\\
     \textbf{Draco II}	 & Pal 4     & - & 2 & - & - & \textbf{Ursa Minor I} &Pal 4	     & 2 & 2 & 2 & 2 \\
     & & & & & & & & & & &\\

	\hline    
\end{tabular}
\caption{The MW GCs that are likely to be associated with a DSG, in H1, H3, L1, and L3 models. Flag 1 and Flag 2 correspond to high and low probabilities of association, respectively.}
\label{tab:DSGs_GCs}
\end{table*}

\subsection{GCs associated with the LMC and the SMC }\label{sec:GC_asso_LMC}
\figref{fig:OrbitalElements_LMC} shows the distribution of the LMC runaway GCs in H1 and H2 models. Similar to the Sgr, the dichotomous pattern can be seen in the figure. In particular, a population of GCs can be seen in the range of $[30,80]\kpc$. These GCs have a larger orbital inclination compared to the GCs at larger distances.  Moreover, compared to H1 model, the distribution of the orbital parameters of GCs expands more in H2 model in which the dynamical friction is considered in the orbital history of the LMC, causing the LMC orbit to spiral inwards.

We should point out that the main caveat of our approach is that the LMC and SMC are assumed to be in an equilibrium state by adopting a simplified Plummer density profile, which does not necessarily match the observed profiles. In order to show the sensitivity of the results to the adopted density profile, new calculations were performed based on the Hernquist model for the density profile while both mass and half-mass radius remained consistent with \citet{bekki2009}. As illustrated in \figref{fig:Plum-Hern}, there is not a significant difference in the orbital parameters of runaway GCs. Therefore, we conclude that the total mass of a DSG is the most effective parameter in the distribution of its runaway GCs rather than the density profile of the host DSGs.

\figref{fig:LMC_Orbit} shows the LMC orbit and two runaway GCs in H1 model. This figure implies that there could be GCs that have escaped from the LMC and are likely to be found at smaller distances from the MW. In the same manner, there could be runaway GCs located at very large distances from the MW. It is clear from \figref{fig:OrbitalElements_LMC} that despite the wide distribution of orbital parameters of runaway GCs, their maximum orbital inclination is less than $38\degr$. This condition is a strong constraint, leading to only a few MW GCs meeting the criteria for association with the LMC. For the H1 model, NGC 5024, Pyxis, and Pal 3 belong to Flag 1, and Pal 4, NGC 7492, and NGC 5053 belong to Flag 2. For the H2 model, NGC 5024, Pyxis, Pal 3, and Pal 4 belong to Flag 1, and NGC 5053 with NGC 7492 belong to Flag 2. According to \citet{massari2019} the Pyxis and Pal 3 are considered to belong to high-energy group GCs.

\par The left panel of \figref{fig:OrbitalElements_SMC} shows the SMC runaway GCs in the H2 model. GCs are distributed in the range of $[50, 180]\kpc$ and $e=[0.0,0.4]$. The maximum orbital inclination of GCs from the SMC orbit is about $10\degr$. With these conditions, there are not any MW GCs associated with the SMC. This is in contrast to \citetalias{Rostami2022} which predicted that two GCs are likely to be associated with the SMC.
\par Owing to its large mass, the LMC can in principle affect the distribution of runaway GCs from DSGs close to the LMC. The distance and inclination of the SMC from the LMC are small, so the presence of the LMC can have a great impact on the distribution of GCs escaping the SMC. The right panel of \figref{fig:OrbitalElements_SMC}, shows the orbital parameters of the SMC runaway GCs in the H2 model and in the presence of the LMC. As evident from the figure, the LMC changes the distribution of runaway GCs significantly leading to the orbital inclination of runaway GCs reaching $\sim60\degr$. Interestingly, a number of runaway SMC GCs are captured by the LMC. If we increase the LMC mass to $\Msun[11]$ \citep{Erkal2019}, more GCs will be captured by the LMC. Since the distribution of runaway SMC GCs is strongly dependent on the mass and presence of the LMC, the association of MW GCs with the SMC alone, cannot be studied in the context of this study. 

\subsection{GCs associated with other MW DSGs}\label{sec:GC_asso_DSG}
As explained in \secref{sec:prob_DSG} we use the Sgr filters for other DSGs as well, except for the LMC and the SMC. In particular, we consider the Sgr filters for H1, H3, L1, and L3 models. For the orbital eccentricity and inclination, the filter values are the same for all models, i.e. $\FSgrE=0.15$ and $\FSgrTheta=30\degr$. However, for the apogalactic distance, we have $\FSgrR=0.8$ in L1 and L3 and $\FSgrR=0.6$ for H1 and H3 models. 

\par Using these filters we identify 19 MW GCs which are associated with a DSG, of which eight GCs belong to Flag 1, and 11 GCs to Flag 2. The results are summarized in \tabref{tab:DSGs_GCs}. The table shows that there exists more than one ex-situ origin for a number of GCs. As a result, there are GCs for which we cannot reliably pinpoint only one DSG for their origin. We refer to these cases as multiple identifications. However, the category of association likelihoods, i.e. flag numbers, can rectify this issue to some extent.

\par \citetalias{Rostami2022} estimated that about two GCs should have escaped from the Fornax. In the present study, we only found one GC with Flag 2, namely Crater, which was associated with the Fornax in all simulation models. \citet{Mucciarelli} investigated chemical abundances of NGC 2005 in the LMC. They concluded that it may have escaped from the Fornax and was later been captured by the LMC. As a result, Crater and NGC 2005 could be the two GCs that we expect to have escaped from the Fornax.

\citet{massari2019} identified a population of eleven loosely bound GCs (AM 1, Eridanus, Pyxis, Palomar 3, Palomar 4, Crater, NGC 6426, NGC 5694, NGC 6584, NGC 6934 and Palomar 14) in the MW that do not seem to have formed in-situ. We have found possible ex-situ origins for Eridanus, Pyxis, Palomar 3, Palomar 4, Crater and NGC 6426.

To identify the possible progenitors of MW GCs, \citet{Boldrini2022} performed a set of comprehensive orbit integrations to track 170 GCs and 11 MW DSGs backwards in time in a combined potential of the Milky-Way and satellites. To evaluate possible past associations, they proposed a globular-cluster–satellite binding criterion based on the satellite’s tidal radius and escape velocity. They found that only 6 GCs were effectively associated with the Sgr. NGC 6715, Arp 2, Terzan 7 and Terzan 8 still belong to Sgr, whereas Whiting 1 and Palomar 12 were accreted by the MW less than $0.3\Gyrs$ ago. According to their results, the other 164 GCs, have not been associated with MW DSGs. The difference between our results and their conclusion could be due to the binding criteria that is used in \citet{Boldrini2022}. They assumed that the orbital radius of a GC centered on a galaxy at a specific time ${r_\mathrm{c}}^\mathrm{GC} (t_\mathrm{i})$ to be smaller than the satellite tidal radii at $t_\mathrm{i}$, and the GC relative velocity with respect to its putative parent galaxy to be lower than the escape velocity of the satellite at $t_\mathrm{i}$. It should be noted that in the binding criterion method of \citet{Boldrini2022}, ${r_\mathrm{c}}^\mathrm{GC} (t)$ parameter is highly sensitive to the assumed model of MW-mass growth as well as DSGs mass-loss histories and dynamical friction, while the orbital parameter space for runaway GCs is not sensitive to the initial conditions of the host DSGs and MW models. It is expected that the binding criterion method \citep{Boldrini2022}, will give correct results in short periods of time ($0.5\Gyr$). It will be important to continue testing both methods with Gaia DR4 and further full-scale \Nbody modeling of GCs evolving within a dynamic triaxial galactic potential of DSGs and the MW to reach a consensus. 

\subsection{Runaway GCs from dissolved DSGs}\label{sec:GC_merg}
\begin{figure}
  \centering
  \includegraphics[width=1\linewidth]{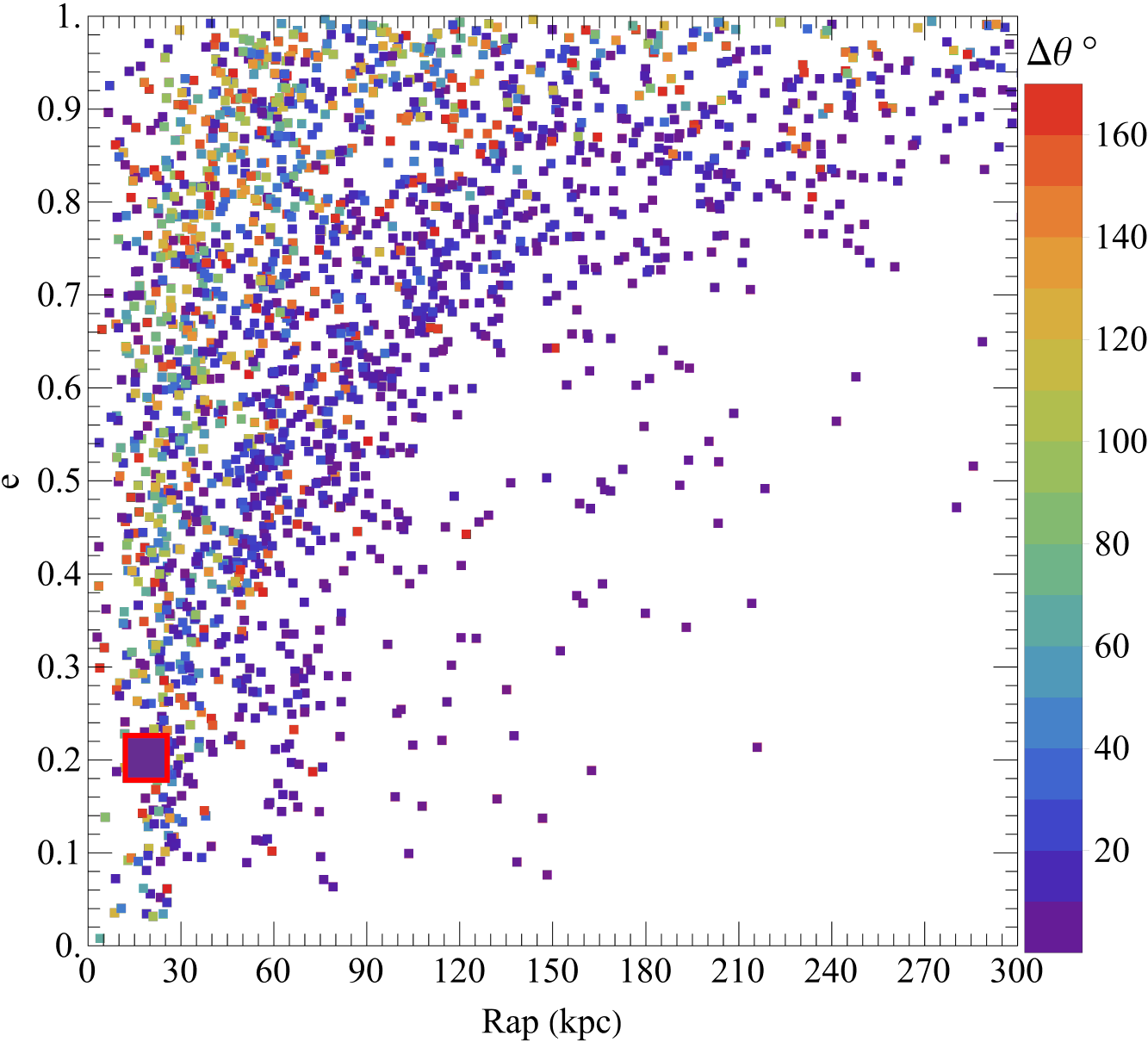}
  \caption{The orbital parameters of 2500 GCs escaped from a dissolved DSG in the H3 model after $12\Gyrs$. The DSG was placed on the MW disc with an orbital eccentricity of $e=0.2$ and an apogalactic distance of $\Rap=24
  \kpc$. The DSG was dissolved after $6\Gyrs$. The orbital parameters of the DSG at the time of dissolution are shown by the large square. The colour coding indicates the orbital inclination of these GCs with respect to the DSG orbit (in degrees).}
  \label{fig:OrbitalElements_Dissolved_DSG}
\end{figure}

\begin{figure*}
	\centering
	\includegraphics[width=0.9\linewidth]{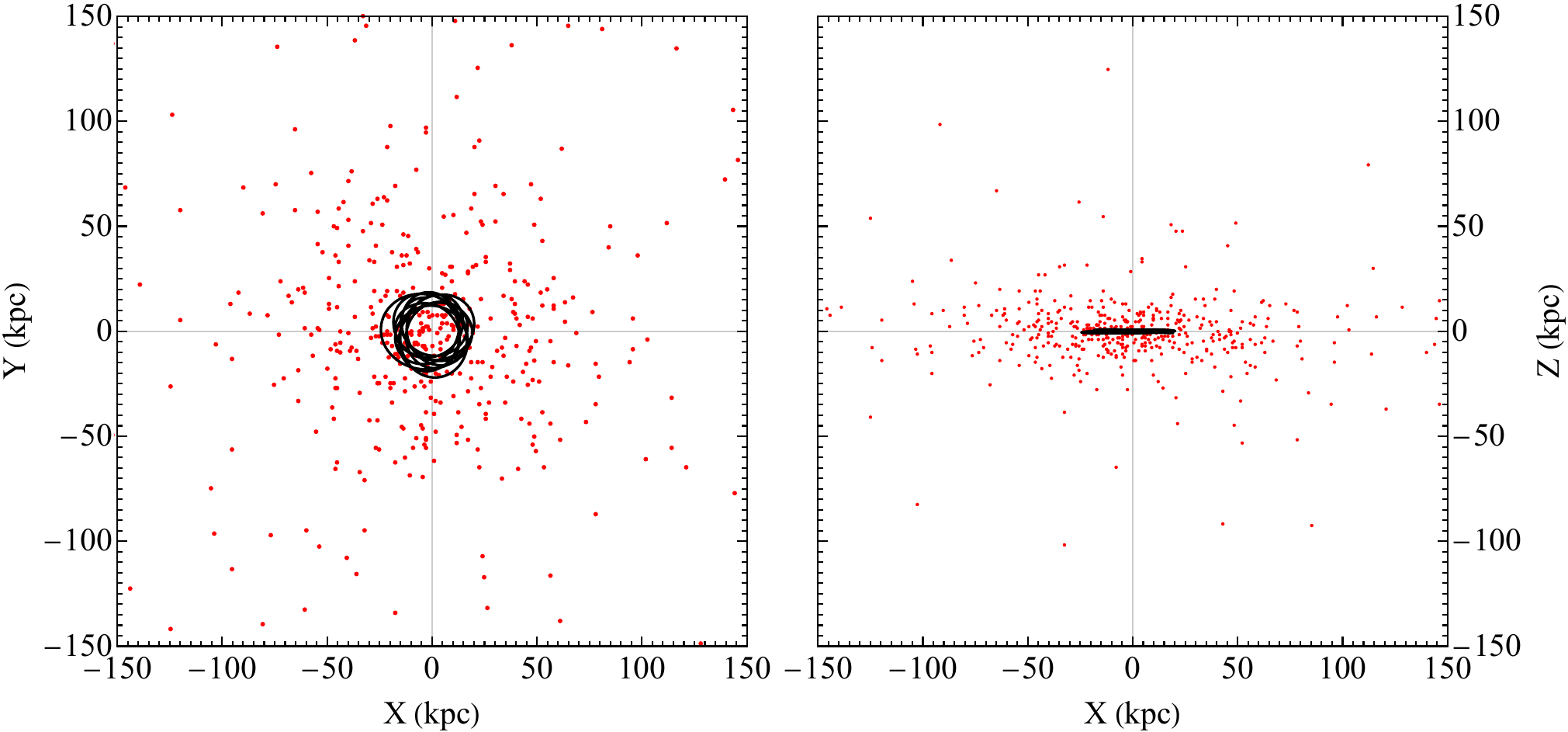}  	
	\caption{The spatial distribution of 500 runaway GCs (red dots) from a dissolved DSG in the H3 model. The DSG has been dissolved in the first $6\Gyrs$ and its orbit is shown in black.}  
	\label{fig:Dissolved_DSG}
\end{figure*}

As mentioned in \secref{sec:intro}, we expect that a number of MW GCs come from DSGs that have been completely merged with the MW. Such DSGs must have two important characteristics. First, they must be massive enough to form GCs, and second, they must be close enough to the MW to merge with it completely despite their large masses. 

\par In this section we study the distribution of orbital parameters of runaway GCs from a dissolved DSG. For this purpose, as we did for the Sgr in the H3 models, we must first solve the \Nbody problem to obtain the potential field of the DSG as a function of position and time, i.e. $\phi(\vec{r},t)$. We perform the simulation from $t_0=-12\Gyrs$ until the present time. As described in \secref{sec:sim_model} for the Sgr, we consider 50,000 particles of the same mass distributed according to the Plummer model with a total mass of $5\times\Msun[10]$ and a half-mass radius of 4$\kpc$. We place the DSG on the MW disc with an orbit of $\Rap=24\kpc$ and $e=0.2$. As the MW potential grows stronger, the eccentricity of the DSG orbit does not change significantly, whereas $\Rap$ reaches $17\kpc$ at the time of its dissolution, which is $T\approx6\Gyrs$. Now that $\phi(\vec{r},t)$ is known, we proceed to determine the orbital parameters of runaway GCs at the present time. \figref{fig:OrbitalElements_Dissolved_DSG} shows the distribution of orbital parameters of these GCs.     
\par As shown in the figure, the orbits of escaped GCs cover a very wide range and there is almost no correlation between the orbital parameters of these GCs and the DSG. Due to the strong gravitational force of the DSG and its proximity to the MW, the orbits of GCs get 
severely disturbed, leading them to be thrown in different directions. Another interesting point is that there are runaway GCs with very large apogalactic distances. This indicates that even GCs that are very far from the MW, could come from a dissolved DSG. In \secref{sec:eff_mass} we showed that for light DSGs, the distribution of orbital parameters of GCs covers a smaller range. However, it should be noted that the initial mass of these DSGs can not be considered too low, since they can not form GCs. The wide distribution of GCs orbits makes it very difficult to find their origin (host DSG). As an example, the origin of a GC with $\Rap=200\kpc$ and $i=70\degr$, can be the same as that of a GC located on the MW disc at a distance of $\Rap=8\kpc$. \figref{fig:Dissolved_DSG} shows the present-day spatial distribution of 500 GCs escaped from the dissolved DSG. As evident from the figure, the runaway GCs are widely distributed in space. For such GCs, complementary photometric and spectroscopic observations can help in identifying their possible ex-situ origin.

\par Using the data from \Gaia DR2 and the proper motions from the Hubble Space Telescope, \citet{Myeong1018} studied 91 MW GCs in energy-action space, to investigate their possible association with the \Gaia-Enceladus-Sausage structure. They showed that eight of the high-energy, old halo GCs are strongly clustered in both vertical and azimuthal axes, but are well spread out radially. In addition, the GCs exhibit a large radial anisotropy with highly eccentric orbits, i.e. $e>0.8$. Their apogalactic distances and orbital inclination range from $12$ to $20\kpc$, and $5$ to $170\degr$, respectively. They concluded that the characteristics of these GCs are consistent with a DSG merged with the MW, which could be the origin of the \Gaia-Enceladus-Sausage. They estimated an initial mass of $\sim5\times\Msun[10]$ for the DSG. We argue that their compiled list of candidates could be incomplete. We demonstrated in \figref{fig:Dissolved_DSG} that for a dissolved DSG, the runaway GCs occupy a wide area in the parameter space. For example in \figref{fig:Dissolved_DSG}, the eccentricity of the runaway GCs can be any value from 0.0 to 1.0. This is in contrast to the results of \citet{Myeong1018}, where all their candidates have $e>0.8$. This implies that they could be, in principle, excluding some GCs. However, due to different methods and selection criteria, we cannot readily compare their results with ours and further elaborate on this discrepancy. This could be the subject of a future study, where we exclusively investigate the association of the MW GCs with merging processes whose progenitor DSGs are completely dissolved.

\begin{table*}
\begin{tabular}{cccccccccccc}
    \hline
    & & & & & & \\
	DSG & $\alpha$ & $\delta$ & $\Dh$ & $\mualphac$ & $\mudelta$ & $\VLOS$\\
	 & (deg) & (deg) & $(\kpc)$ & $(\masyr)$ & $(\masyr)$ & $(\kms)$ \\
	\hline 
    Antlia II     & $143.88^{[5]}$  & $-36.76^{[5]}$  & $132^{[5]}$   & $-0.095\pm0.018^{[5]}$ & $ 0.058\pm0.024^{[5]}$ & $  290.7\pm0.5^{[5]}$ \\ 
    Aquarius II     & $338.48^{[8]}$  & $-9.32^{[8]}$  & $107.9^{[8]}$   & $-0.252\pm0.526^{[1]}$ & $ 0.011\pm0.448^{[1]}$ & $  -71.1\pm2.5^{[1]}$ \\ 
    Boötes I     & $210.02^{[4]}$  & $14.50^{[4]}$  & $66^{[4]}$   & $-0.554\pm0.092^{[1]}$ & $-1.111\pm0.068^{[1]}$ & $  99\pm2.1^{[1]}$ \\ 
    Boötes II     & $209.52^{[4]}$  & $12.85^{[4]}$  & $42^{[4]}$   & $-2.686\pm0.389^{[1]}$ & $-0.530\pm0.287^{[1]}$ & $ -117\pm5.2^{[1]}$ \\ 
    CanVen I      & $202.01^{[6]}$  & $ 33.55^{[6]}$  & $218^{[2]}$   & $-0.159\pm0.094^{[1]}$ & $-0.067\pm0.054^{[1]}$ & $  30.9\pm0.6^{[1]}$  \\  
    CanVen II     & $194.29^{[6]}$  & $34.32^{[6]}$  & $160^{[2]}$   & $-0.342\pm0.232^{[1]}$ & $-0.473\pm0.169^{[1]}$ & $ -128.9\pm1.2^{[1]}$ \\ 
    Carina I      & $100.40^{[3]}$  & $-50.96^{[3]}$  & $105.2^{[7]}$ & $ 0.485\pm0.017^{[1]}$ & $ 0.131\pm0.016^{[1]}$ & $ 229.1\pm0.1^{[1]}$ \\
    Carina II     & $114.10^{[4]}$  & $-57.99^{[4]}$  & $36.2^{[4]}$   & $1.867\pm0.078^{[1]}$ & $0.082\pm0.072^{[1]}$ & $ 477.2\pm1.2^{[1]}$ \\ 
    Carina III     & $114.63^{[4]}$  & $-57.89^{[4]}$  & $27.8^{[4]}$   & $3.046\pm0.119^{[1]}$ & $1.565\pm0.135^{[1]}$ & $ 284.6\pm3.4^{[1]}$ \\ 
    Coma Berenices     & $186.74^{[4]}$  & $23.90^{[4]}$  & $42^{[4]}$   & $0.471\pm0.108^{[1]}$ & $-1.716\pm0.104^{[1]}$ & $98.1\pm0.9^{[1]}$ \\  
    Crater II     & $177.31^{[6]}$  & $-18.41^{[6]}$  & $116.5^{[9]}$   & $-0.184\pm0.061^{[1]}$ & $-0.106\pm0.031^{[1]}$ & $87.5\pm0.4^{[1]}$ \\     
    Draco I       & $260.05^{[3]}$  & $ 57.91^{[3]}$  & $ 75.9^{[7]}$ & $-0.012\pm0.013^{[1]}$ & $-0.158\pm0.015^{[1]}$ & $-291\pm0.1^{[1]}$  \\  
    Draco II     & $238.19^{[4]}$  & $64.56^{[4]}$  & $21.5^{[4]}$   & $1.242\pm0.276^{[1]}$ & $0.845\pm0.285^{[1]}$ & $-347.6\pm1.8^{[1]}$ \\ 
    Eridanus II     & $56.08^{[6]}$  & $-43.53^{[6]}$  & $366^{[10]}$   & $0.159\pm0.292^{[1]}$ & $0.372\pm0.34^{[1]}$ & $75.6\pm2.4^{[1]}$ \\   
    Fornax I      & $ 39.99^{[3]}$  & $-34.44^{[3]}$  & $147.2^{[7]}$ & $ 0.374\pm0.004^{[1]}$ & $-0.401\pm0.005^{[1]}$ & $  55.3\pm0.3^{[1]}$ \\
    Grus I     & $344.17^{[6]}$  & $-50.16^{[6]}$  & $120^{[11]}$   & $-0.261\pm0.172^{[1]}$ & $-0.437\pm0.238^{[1]}$ & $-140.5\pm2.4^{[1]}$ \\ 
    Hercules I     & $247.75^{[6]}$  & $12.79^{[6]}$  & $132^{[2]}$   & $-0.297\pm0.118^{[1]}$ & $-0.329\pm0.094^{[1]}$ & $45\pm1.1^{[1]}$ \\
    Horologium I     & $43.88^{[4]}$  & $-54.11^{[4]}$  & $87^{[4]}$   & $0.891\pm0.088^{[1]}$ & $-0.550\pm0.080^{[1]}$ & $112.8\pm2.6^{[1]}$ \\
    Hydra II     & $185.42^{[6]}$  & $-31.98^{[6]}$  & $134^{[12]}$   & $-0.416\pm0.519^{[1]}$ & $0.134\pm0.422^{[1]}$ & $303.1\pm1.4^{[1]}$ \\
    Hydrus I     & $37.38^{[4]}$  & $-79.30^{[4]}$  & $27.6^{[4]}$   & $3.733\pm0.038^{[1]}$ & $-1.605\pm0.036^{[1]}$ & $80.4\pm0.6^{[1]}$ \\    
    Leo I         & $152.11^{[3]}$  & $ 12.30^{[3]}$  & $253.5^{[7]}$ & $-0.086\pm0.059^{[1]}$ & $-0.128\pm0.062^{[1]}$ & $ 282.5\pm0.5^{[1]}$ \\ 
    Leo II     & $168.37^{[6]}$  & $22.15^{[6]}$  & $233.30^{[7]}$   & $-0.025\pm0.080^{[1]}$ & $-0.173\pm0.083^{[1]}$ & $78.0\pm0.1^{[1]}$ \\
    Leo IV     & $173.23^{[6]}$  & $-0.53^{[6]}$  & $154^{[2]}$   & $-0.590\pm0.531^{[1]}$ & $-0.449\pm0.358^{[1]}$ & $132.3\pm1.4^{[1]}$ \\
    Leo V      & $172.79^{[6]}$  & $2.22^{[6]}$   & $178^{[2]}$   & $-0.097\pm0.557^{[1]}$ & $-0.628\pm0.302^{[1]}$ & $173.3\pm3.1^{[1]}$ \\
    Phoenix I     & $27.77^{[6]}$  & $-44.44^{[6]}$  & $418^{[13]}$   & $0.079\pm0.099^{[1]}$ & $-0.049\pm0.120^{[1]}$ & $-21.2\pm1.0^{[1]}$ \\
    Pisces II     & $344.62^{[6]}$  & $5.95^{[6]}$  & $180^{[14]}$   & $-0.108\pm0.645^{[1]}$ & $-0.586\pm0.498^{[1]}$ & $-226.5\pm2.7^{[1]}$ \\
    Reticulum II     & $53.92^{[4]}$  & $-54.04^{[4]}$  & $32^{[4]}$   & $2.398\pm0.040^{[1]}$ & $-1.319\pm0.048^{[1]}$ & $62.8\pm0.5^{[1]}$ \\    
    Sculptor I    & $ 15.03^{[3]}$  & $-33.70^{[3]}$  & $ 85.9^{[7]}$ & $ 0.084\pm0.006^{[1]}$ & $-0.133\pm0.006^{[1]}$ & $ 111.4\pm0.1^{[1]}$ \\
    Segue 1     & $151.76^{[4]}$  & $16.08^{[4]}$  & $23^{[4]}$   & $-1.697\pm0.195^{[1]}$ & $-3.501\pm0.175^{[1]}$ & $208.5\pm0.9^{[1]}$ \\ 
    Segue 2     & $34.81^{[4]}$   & $20.17^{[4]}$  & $35^{[4]}$   & $1.656\pm0.155^{[1]}$ & $0.135\pm0.104^{[1]}$ & $-39.2\pm2.5^{[1]}$ \\    
     Sextans I     & $153.26^{[3]}$  & $ -1.61^{[3]}$  & $ 85.9^{[7]}$ & $-0.438\pm0.028^{[1]}$ & $ 0.055\pm0.028^{[1]}$ & $ 224.2\pm0.1^{[1]}$ \\ 
    Triangulum II     & $33.32^{[4]}$   & $36.17^{[4]}$  & $30^{[4]}$   & $0.588\pm0.187^{[1]}$ & $0.554\pm0.161^{[1]}$ & $-381.7\pm1.1^{[1]}$ \\ 
    Tucana II     & $342.97^{[4]}$   & $-58.56^{[4]}$  & $58^{[4]}$   & $0.910\pm0.059^{[1]}$ & $-1.159\pm0.074^{[1]}$ & $-129.1\pm3.5^{[1]}$ \\ 
    Tucana III     & $359.15^{[4]}$   & $-59.60^{[4]}$  & $25^{[4]}$   & $-0.025\pm0.034^{[1]}$ & $-1.661\pm0.035^{[1]}$ & $-102.3\pm2^{[1]}$ \\    
    Ursa Major I  & $158.72^{[4]}$  & $ 51.92^{[4]}$  & $ 97.3^{[4]}$ & $-0.683\pm0.094^{[1]}$ & $-0.72\pm0.13^{[1]}$  & $ -55.3\pm1.4^{[1]}$ \\
    Ursa Major II     & $132.87^{[4]}$   & $63.13^{[4]}$  & $34.7^{[4]}$   & $1.691\pm0.053^{[1]}$ & $-1.902\pm0.066^{[1]}$ & $-116.5\pm1.9^{[1]}$ \\    
    Ursa Minor I  & $227.28^{[3]}$  & $ 67.22^{[3]}$  & $ 75.9^{[7]}$ & $-0.184\pm0.026^{[1]}$ & $ 0.082\pm0.023^{[1]}$ & $-246.9\pm0.1^{[1]}$  \\
    Willman 1     & $162.33^{[4]}$   & $51.05^{[4]}$  & $45^{[4]}$   & $0.199\pm0.187^{[1]}$ & $-1.342\pm0.366^{[1]}$ & $-12.3\pm2.5^{[1]}$ \\ 
    Sgr I         & $283.83^{[3]}$  & $-30.54^{[3]}$  & $ 26.3^{[7]}$ & $-2.736\pm0.009^{[1]}$ & $-1.357\pm0.008^{[1]}$ & $ 140\pm 2^{[1]}$ \\ 
    SMC           & $12.80^{[3]}$ & $-73.15^{[3]}$ & $ 64^{[7]}$   & $ 0.797\pm0.03^{[3]}$ & $-1.220\pm0.03^{[3]}$ & $ 145.6\pm0.6^{[3]}$ \\ 
    LMC           & $81.28^{[3]}$ & $-69.78^{[3]}$ & $ 50.6^{[7]}$ & $ 1.850\pm0.03^{[3]}$ & $ 0.234\pm0.03^{[3]}$ & $ 262.2\pm3.4^{[3]}$  \\
    \hline
    
\end{tabular}    
\caption{Astrometric data of 41 of the most massive MW DSGs. Columns 2 and 3 represent the equatorial coordinates $(\alpha, \delta)$. The heliocentric distance of each DSG is denoted by $\Dh$ and is given in column 4. Columns 5 and 6 show the components of proper motions in the direction of right ascension ($\mualphac$) and declination ($\mudelta$), respectively. The last column represents the line-of-sight velocity ($V_\mathrm{LOS}$). \textit{References}: (1) \citet{Fritz2018}, (2) \citet{Alan2012}, (3) \citet{Helmi2018}, (4) \citet{Simon2018}, (5) \citet{Torrealba2019}, (6) \citet{Alan2020}, (7) \citet{Pawlowski2013}, (8) \citet{Torrealba2016}, (9) \citet{Vivas2020}, (10) \citet{Crnojevi2016}, (11) \citet{Walker2016}, (12) \citet{Martin2015}, (13) \citet{Siegert2016}, (14) \citet{Belokurov2010}}
\label{tab:MW_DSGs_params}
\end{table*}

\section{Conclusion}\label{sec:conclusion}
As a follow-up to \citetalias{Rostami2022}, we carried out a large number of three-body simulations to obtain the distribution of possible runaway GCs from the MW DSGs. To obtain the distribution we assumed GCs as point masses and DSGs as a Plummer model (\secref{sec:sgrpot}). For the MW we considered two sets of static and dynamic models which account for the sustained growth of the MW since its birth (\secref{sec:sim_model}). The MW models are constituted by three components, namely a bulge with a power-law profile and an exponential cutoff, a \citet{Miyamoto1975} model for the disc, and an NFW dark-matter halo (\secref{sec:MWpot}). We took the effect of dynamical friction into account as well. We used a Runge-Kutta integrator of the 8th order over an interval of $8\Gyrs$.    

\par We then compared the distribution of the runaway GCs from each DSG with that of the MW GCs in the space of orbital parameters, namely $(\Rap, e, \Delta\theta\degr)$, where $\Rap(\kpc)$ is the apogalactic distance, $e$ is eccentricity, and $\Delta\theta\degr$ is the orbital inclination of GCs with respect to the orbital plane of their host DSG. We quantified the association likelihood for each GC-DSG pair by assigning them an association flag (\secref{sec:prob_sgr}). The Flag 1 category represents a high association probability, whereas Flag 2 corresponds to a lower association probability. The main outcomes of the study are the following.

\begin{itemize}
    \item A comparison between the results of our three-body method and that of the \textsc{NBODY6} code, demonstrates that the runaway particles exhibit the same distribution in the orbital parameter space and the observed trends with respect to e.g. the DSG mass are the same.
    
    \item We observed that the distribution of runaway GCs is not random and does not cover the entire parameter space. Rather, it exhibits a dichotomous pattern. In particular, the runaway GCs form two distinctive populations, a population in which  $\Rap(\mathrm{GC})<\Rap(\mathrm{DSG})$ and another one with $\Rap(\mathrm{GC})>\Rap(\mathrm{DSG})$. We further proved the existence of such a dichotomy using a semi-analytical approach. Moreover, the values of $\Rap$ and $e$ for runaway GCs are positively correlated with each other. In other words, GCs with larger (smaller) values of $\Rap$ are expected to have larger (smaller) values of $e$ on average or vice versa.
    
    \item There exists a positive correlation between the mass of a DSG and the dispersion of its runaway GCs in the parameter space. This implies that for massive DSGs such as the LMC, an associated runaway GC could be as far as $\Rap=400\kpc$ or as close as $\Rap=30\kpc$. In contrast, the values of $\Rap$ and $e$ for runaway GCs of a light DSG such as Carina I are expected to lie within a very small range from $\Rap$ and $e$ of Carina I. This correlation renders the classical methods of identifying associations between MW GCs and DSGs error-prone. In other words, one cannot simply look for clusters in the parameter space and associate them with a neighboring DSG. For massive DSGs there could be a large tail of GCs in the parameter space which are not clustered around the DSG and their association will be probably ruled out. 
      
    \item Related to the previous point, the dispersion in the parameter space for the runaway GCs of dissolved DSGs is even larger. The runaway GCs occupy a large fraction of the parameter space for dissolved DSGs. For a dissolved DSG with ($\Rap=24\kpc$, $e=0.2$), the apogalactic distances of runaway GCs can be any value from $\Rap=5$ to $\Rap=300\kpc$ and their orbital eccentricities can range from $e=0.0$ to $e>0.9$. The only statement which can be made about such DSGs is that their runaway GCs still exhibit the aforementioned positive correlation between $\Rap$ and $e$. Apart from this, there are not any other patterns or correlations between the distribution of runaway GCs and their dissolved host DSG. This makes it extremely difficult, if not entirely impossible, to identify associations for dissolved DSGs, within the context of purely kinematic studies like ours.
    
    \item For DSGs that are close to massive DSGs such as the Sgr or the LMC, the orbital parameters of runaway GCs can be significantly perturbed by the massive DSG, to the extent that, some of the GCs might be even recaptured by the massive DSG. This is why for the SMC, we needed to take the presence of the LMC into account.
    \item We identified 18 GCs which are associated with the Sgr. Combined with the fact that the Sgr still has four bound GCs, this is in agreement with \citetalias{Rostami2022} where we anticipated about 14 GCs should have escaped from the Sgr so far. According to a number of studies the following five GCs have been suggested to have originated from the Sgr, NGC~6284, NGC~4147, NGC~5053, NGC~5824, and Pal~2. However, we could not confirm their kinematic association. We report a number of new associations with the Sgr for the first time. They are NGC~5466, NGC~4590, Rup~106, and Pal~4, all which have a high probability of association. Moreover, we identified NGC~6101, NGC~5897, NGC~6235, NGC~6934, NGC~6426, and IC~4499, albeit with a lower probability. Other associations we identified are Whiting~1, NGC~6715 (M~54), Ter~7, Arp~2, Ter~8, Pal~12, NGC~2419, and NGC~5634, all of which are in agreement with other studies.    
    \item In \citetalias{Rostami2022} we predicted four possible associations with the LMC. Here, for the H1 model, the high probability associations are NGC 5024, Pyxis, and Pal~3, and lower probability ones are Pal 4, NGC 7492, and NGC 5053. For the H2 model, the results are the same as H1, except that Pal~4 is added to Flag 1 associations as well. Intriguingly, \citet{massari2019} has categorized Pal~3 and Pyxis in the high-energy group and their origin as uncertain.   
    \item According to \citetalias{Rostami2022}, we expect two runaway GCs from the SMC. In the present paper we demonstrated that the LMC has a non-negligible influence on the dispersion of the SMC runaway GCs. In particular, we observed that the SMC runaway GCs could be recaptured by the LMC. As a result, we cannot conclusively constrain the SMC associations in the context of our current study.    
    \item For the rest of the MW DSGs, which have lower masses, we designated 19 associations, of which eight are of Flag~1, and 11 of Flag~2.  For the Fornax, we had predicted two runaway GCs \citepalias{Rostami2022}. We found Crater as a possible association with the Fornax. Furthermore, another Fornax runaway GC might now reside in the LMC \citep{Mucciarelli}. This is consistent with our findings for the Fornax.
    \item In total, we identified 29 MW GCs which could have originated from DSGs. This indicates that a maximum of $19\percent$ of all MW GCs could have an ex-situ origin.    
    \item For a number of GCs, we found several ex-situ origins. We refer to these cases as multiple identifications. Our categories of association likelihood, aka flags, can rectify this issue to some extent but not completely. As a result, there are GCs for which we cannot reliably pinpoint only one DSG and their origin still remains an open question.  
    \item Finally, we find a concentration of simulated runaway GCs from the Sgr which are clustered around $\Rap\approx275\kpc$ for the L4 model, and $\Rap\approx375\kpc$ for the H4 model, both of which have $e\approx0.8$ and $\Delta\theta\approx20\degr$. These values correspond to the orbital conditions of the Sgr at the beginning of the simulations, i.e. at $t=-8\Gyr$, where the Sgr was located at larger distances due to the effect of dynamical friction. So far, there are not any MW GCs observed with such orbital parameters. This could be due to an observational bias since such GCs might not be sufficiently bright to be easily detected. Another possibility is that such GCs have entirely dissolved owing to their low initial mass. This in principle can lead to a dispersed stream with a low surface brightness which also poses an observational challenge. A third option could be that such GCs do not exist at all. This is an intriguing outcome of our study which is worth further (observational) investigations to determine which of the scenarios is the case.
\end{itemize}

There are a number of shortcomings to our approach. First, except for a few cases, we mainly model DSGs with a static and smooth potential, namely the Plummer profile. More robust treatment of DSGs requires utilizing particle-mesh codes such as \textsc{SUPERBOX} \citep{Fellhauer2000}. Such an approach accounts for the internal dynamics of the DSGs, their deformation, tidal stripping,  formation of tails, and their dissolution. However, this is computationally expensive if one attempts to sweep a large parameter space. This is an advantage of our method which allows running a large ensemble of simulations with different initial conditions. Having found the distribution of runaway GCs in the parameter space, one can now perform more representative simulations using e.g. \textsc{SUPERBOX}. This can be the subject of future study.  

Second, our approach has been entirely kinematic, hence providing only one piece of a puzzle, albeit a vital piece. Our findings can be substantially complemented by photometric and spectroscopic studies. A thorough multivariate analysis of all observational data for each GC and DSG is beyond the scope of the present work and is left to a forthcoming paper. A major benefit of such multivariate studies is the possibility to resolve the issue of singularities or multiple identifications. Moreover, one has the means to examine peculiar cases, such as a GC which has been recaptured by another DSG. Such a GC will exhibit photometric/spectroscopic properties which are different from its present-day host but resemble those of another DSG. 

As an interesting research direction for future, one can compare our findings with the outcome of high resolution cosmological simulations, e.g. \textsc{FIRE} \citep{Hopkins2014, Wetzel2023} and \textsc{APOSTLE} \citep{Sawala2016}.

\section*{Data availability}
The data underlying this article are available in the article.

\bibliographystyle{mnras}
\bibliography{article} 
\bsp
\label{lastpage}
\end{document}